\documentstyle[psfig,aps,prd,eqsecnum]{revtex}

\title{Searching for periodic sources with LIGO.  II: Hierarchical
searches}
\author{Patrick R Brady}
\address{Institute for Theoretical Physics, University of
California, Santa Barbara, CA 93106 \\
Department of Physics,  University of Wisconsin-Milwaukee, PO Box 413,
Milwaukee, WI 53201}
\author{Teviet Creighton}
\address{Theoretical Astrophysics, California
Institute of Technology, Pasadena, CA 91125}
\date{3 December 1998}

\begin{document}

\def\ptitle#1{\wideabs
{\maketitle \abstract #1
\endabstract \pacs{PACS numbers: 03.70.+k, 98.80.Cq}}}


\draft

\ptitle{ The detection of quasi-periodic sources of gravitational
waves requires the accumulation of signal-to-noise over long
observation times.  This represents the most difficult data analysis
problem facing experimenters with detectors like those at LIGO.  If
not removed, Earth-motion induced Doppler modulations and intrinsic
variations of the gravitational-wave frequency make the signals
impossible to detect.  These effects can be corrected (removed) using
a parameterized model for the frequency evolution.  In a previous
paper, we introduced such a model and computed the number of
independent parameter space points for which corrections must be
applied to the data stream in a coherent search.  Since this number
increases with the observation time, the sensitivity of a search for
continuous gravitational-wave signals is computationally bound when data
analysis proceeds at a similar rate to data acquisition.  In this
paper, we extend the formalism developed by Brady \textit{et al.}
\mbox{[}Phys. Rev. D {\bf 57}, 2101 (1998)\mbox{]}, and we compute the
number of independent corrections $N_p(\Delta T, N)$ required for
incoherent search strategies.  These strategies rely on the method of
\emph{stacked} power spectra---a demodulated time series is divided
into $N$ segments of length $\Delta T$, each segment is Fourier
transformed, a power spectrum is computed, and the $N$ spectra are
summed up.  This method is incoherent; phase information is lost from
segment to segment.  Nevertheless, power from a signal with fixed
frequency (in the corrected time series) is accumulated in a single
frequency bin, and amplitude signal-to-noise accumulates as $\sim
N^{1/4}$ (assuming the segment length $\Delta T$ is held fixed). We
estimate that the sensitivity of an all-sky search that uses
incoherent stacks is a factor of $2$--$4$ better than achieved using
coherent Fourier transforms; incoherent methods are computationally
efficient at exploring large parameter spaces.  We also consider a
two-stage hierarchical search in which candidate events from a search
using short data segments are followed up in a search using longer
data segments.  This hierarchical strategy yields another $20$--$60\%$
improvement in sensitivity in all-sky (or directed) searches for old
($\geq 1000$yr) slow ($\leq 200$Hz) pulsars, and for young ($\geq
40$yr) fast ($\leq 1000$Hz) pulsars.  Assuming enhanced LIGO detectors
(LIGO-II) and $10^{12}$~flops of effective computing power, we also
examine the sensitivity to sources in three specialized classes. A
limited area search for pulsars in the Galactic core would detect
objects with gravitational ellipticities of $\epsilon\agt5\times
10^{-6}$ at 200~Hz; such limits provide information about the strength
of the crust in neutron stars.  Gravitational waves emitted by the
unstable $r$-modes of newborn neutron stars would be detected out to
distances of $\sim 8$~Mpc, if the $r$-modes saturate at a
dimensionless amplitude of order unity and an optical supernova
provides the position of the source on the sky.  In searches targeting
low-mass x-ray binary systems (in which accretion-driven spin up is
balanced by gravitational-wave spin down), it is important to use
information from electromagnetic observations to determine the orbital
parameters as accurately as possible.  An estimate of the difficulty
of these searches suggests that objects with x-ray fluxes which exceed
$2\times10^{-8}\mathrm{erg\,cm}^{-2}\mathrm{s}^{-1}$ would be detected
using the enhanced interferometers in their broadband configuration.
This puts Sco~X-1 on the verge of detectability in a broad-band
search; in this case amplitude signal-to-noise might be increased by
$\sim 5$--$10$ by operating the interferometer in a signal-recycled,
narrow-band configuration.  Further work is needed to determine the
optimal search strategy when we have limited information about the
frequency evolution of a source in a targetted search.}


\section{Introduction}
\label{s:introduction}

The detection of periodic sources of gravitational waves using the
LIGO, or similar, gravitational-wave detectors, is seemingly the most
straightforward data analysis problem facing gravitational wave
astronomers.  It is also the most computationally intensive; extremely
long observation times will be required to have any chance of
detecting these signals.  Searches for periodic (or quasi-periodic)
sources will be limited primarily by the computational resources
available for data analysis, rather than the duration of the signals
or the lifetime of the instrument.  For this reason, it is of
paramount importance to explore different search strategies and to
determine the optimal approach before the detectors go on line at the
end of the century.  In a previous paper~\cite{Brady_P:1998prd},
hereafter referred to as Paper~I, we presented a detailed discussion
of issues which arise when one searches for these sources in the
detector output.  Using a parameterized model for the expected
gravitational wave signal, we presented a method to determine the
number of independent parameter values which must be sampled in a
search using coherent Fourier transforms (which accumulate the signal
to noise in an optimal fashion).  The results were presented in the
context of single-sky-position directed searches, and all-sky
searches, although the method outlined in Paper~I is applicable to
{\em any} search over a specified region of parameter space.
Livas~\cite{Livas_J:1987}, Jones \cite{Jones_G:1995} and Niebauer {\it
et al.}~\cite{Niebauer_T:1993} have implemented variants of the
coherent search technique without the benefit of the optimization
advocated in Paper~I.

In this paper, we discuss alternative search algorithms which can
better detect quasi-periodic gravitational waves using broadband
detectors.  These algorithms achieve better sensitivities than a
coherent search with equivalent available computational resources.
This improvement is accomplished by combining coherent Fourier
transforms with incoherent addition of power spectra, and by using
hierarchical searches which follow up the candidate detections from a
first pass search.

The most likely sources of quasi-periodic gravitational waves in the
frequency bands of terrestrial interferometric detectors are rapidly
rotating neutron stars.  We use these objects as guides when choosing
the scope of the example searches considered below.  Nevertheless, the
search algorithms are sufficient to detect all sources of continuous
gravitational wave signals provided the frequency is slowly changing.

A rotating neutron star will radiate gravitational waves if its mass
distribution (or mass-current distribution) is not symmetric about its
rotation axis.  Several mechanisms which may produce non-axisymmetric
deformations of a neutron star, and hence lead to gravitational wave
generation, have been discussed in the
literature~\cite{Chandrasekhar_S:1970,Friedman_J:1978,%
Bonazzola_S:1996,Zimmermann_M:1980,Zimmermann_M:1979,%
Wagoner_R:1984}.  A neutron star with non-zero quadrupole moment which
rotates about a principle axis produces gravitational waves at a
frequency equal to twice its rotation frequency.  Equally strong
gravitational waves can be emitted at other frequencies when the
rotation axis is not aligned with a principal axis of the
source~\cite{Bonazzola_S:1996,Galtsov_D:1984}.  If the star precesses,
the gravitational waves will be produced at three frequencies: the
rotation frequency, and the rotation frequency plus and minus the
precession frequency~\cite{Zimmermann_M:1979}.

For concreteness, we consider a model gravitational-wave signal with
one spectral component.  This is not a limitation of our analysis
since the search strategy presented below is inherently broadband; it
can be used to detect sources which emit gravitational waves at any
frequency in the detector pass-band.  Additional knowledge of the
spectral characteristics of a signal might allow us to improve our
sensitivity in the case when multiple spectral components have similar
signal-to-noise ratio.  In such a circumstance, a modified search
algorithm would sum the power from each frequency at which radiation
would be expected.  In a background of Gaussian noise, the sensitivity
would improve as $(\mathrm{number\ of\ spectral\ lines})^{1/4}$ for
only a moderate increase in computational cost.

Finally, we mention several other works which consider searching for
quasi-periodic signals in the output of gravitational wave detectors.
Data from the resonant bar detectors around the world has been used in
searches for periodic sources.  New {\it et al.}~\cite{New_K:1995}
have discussed issues in searching for gravitational waves from
millisecond pulsars.  Krolak~\cite{Krolak_A:1998} and Jaranowski {\it
et al.}~\cite{Jaranowski_P:1998a,Jaranowski_P:1998b} have considered
using matched filtering to extract information about the continuous
wave sources from the data stream.  Finally, work is ongoing in
Potsdam to investigate line-tracking algorithms based on the Hough
transform~\cite{Papa_M:1998}; this technique looks quite promising,
although we must await results on the computational cost and
statistical behavior before we can make a detailed comparison to the
techniques described in this paper.

\subsection{Gravitational waveform}
\label{ss:gravitational-waveform}

The long observation times required to detect continuous sources of
gravitational waves make it necessary to account for changes in the
wave frequency; the physical processes responsible for these changes,
and the associated time scales were discussed in Paper~I.  In
addition, the detector moves with respect to the solar system
barycenter (which we take to be approximately an inertial frame),
introducing Doppler modulations of the gravitational wave frequency.
To account for these two effects, we introduce a parameterized model
for the gravitational wave frequency
$f(t;\mbox{\boldmath${\lambda}$\unboldmath})$ and phase
$\phi(t;\mbox{\boldmath${\lambda}$\unboldmath})=2\pi\int\!
f(t;\mbox{\boldmath${\lambda}$\unboldmath}) \,dt$ measured at the
detector:
\begin{eqnarray}
\label{eq:gw-frequency}
f(t;\mbox{\boldmath${\lambda}$\unboldmath}) \! &=& \! f_0
	\left(1+\frac{\vec v}{c}\cdot\hat{n}\right)
	\left(1+\! \sum_{k=1}f_k
	\left[t+\frac{\vec x}{c}\cdot\hat{n}\right]^k
	\right) , \\
\label{eq:gw-phase}
\phi(t;\mbox{\boldmath${\lambda}$\unboldmath}) \! &=& \! 2\pi f_0
	\left(t+\frac{\vec x}{c}\cdot\hat{n}
	+\! \sum_{k=1}\frac{f_k}{k+1}\!
	\left[t+\frac{\vec x}{c}\cdot\hat{n}\right]^{k+1}
	\right) .
\end{eqnarray}
Here $f_0$ is the initial, intrinsic gravitational-wave frequency,
$\vec{x}(t)$ is the detector position, $\vec{v}(t)$ is the detector
velocity, $\hat{n}$ is a unit vector in the direction of the source,
and $f_k$ are arbitrary coefficients which we call \emph{spindown
parameters}.  (We refer the reader to Paper~I for a detailed
discussion of this model and its physical origin.)  The vector
$\mbox{\boldmath${\lambda}$\unboldmath}$ denotes the \emph{search
parameters} --- the parameters of the frequency model that are
(generally) unknown in advance.  In the most general case that we
consider below, the search parameters include frequency $f_0$, the
polar angles $(\theta,\varphi)$ used to specify $\hat{n}$, and the
spindown parameters $f_k$:
\begin{equation}
\label{eq:lambda}
\mbox{\boldmath${\lambda}$\unboldmath} =
(\lambda^0,\lambda^1,\lambda^2,\lambda^3,\lambda^4,\ldots)
	= (f_0,\theta,\varphi,f_1,f_2,\ldots) \; .
\end{equation}
We note that the parameter $\lambda^0=f_0$ defines an overall
frequency scale, whereas the remaining parameters define the shape of
the phase evolution.  It is convenient to introduce the projected
vector $\vec\lambda=(\lambda^1,\lambda^2,\lambda^3,\lambda^4,\ldots)$
of shape parameters alone.

The strain measured at the interferometer is a linear combination of
the $+$ and $\times$ polarizations of the gravitational waves, and
can be written as
\begin{equation}
\label{eq:strain}
h(t;\mbox{\boldmath${\lambda}$\unboldmath}) =
\mathrm{Re}[{\cal A}
e^{-i[\phi(t;\mbox{\scriptsize\boldmath${\scriptsize\lambda}$\unboldmath}) + \Psi]}] \; .
\end{equation}
The time-dependent amplitude ${\cal A}$ and phase $\Psi$
depend on the detector response functions and the orientation of
the source; they vary gradually over the course of a day (see
references \cite{Bonazzola_S:1996,Jaranowski_P:1998a}).  In what follows,
we treat $\cal A$ and $\Psi$ as constants.
Our analysis may be generalized to include the additional phase
modulation; however, this effectively increases the dimension of
the parameter space by one and the number of points that
must be sampled by $\sim 4$, which translates into a reduction in
relative sensitivity of $\sim 6\%$.

\subsection{Parameter ranges}
\label{ss:parameter-ranges}

The computational difficulty of a search for quasi-periodic signals
depends on the range of parameter values that are considered in
the search.  The intrinsic gravitational wave frequency $f_0$
ranges from (near) zero to some cutoff frequency $f_{\mathrm{max}}$.
If gravitational waves are emitted at twice the rotation frequency,
theoretical estimates~\cite{Shapiro_S:1984,Friedman_J:1984} suggest that
\begin{equation}
	f_{\mathrm{max}} \alt 1.2\> \textrm{kHz\ to\ }4\>\textrm{kHz}
\end{equation}
depending on the equation of state adopted in the neutron star model.
Observational evidence---the coincidence of the periods of PSR 1937+21
and PSR 1957+20---favors the lower bound on gravitational wave
frequency $f_{\mathrm{max}} \simeq 1.2\>
\textrm{kHz}$~\cite{Kulkarni_S:1992}.  The spindown parameters $f_j$
are allowed to take any value in the range $|f_j|\leq
(1/\tau_{\mathrm{min}})^j$ where $\tau_{\mathrm{min}}\sim f/\dot{f}$
is the characteristic time scale over which the frequency might be
expected to change by a factor of order unity.  Observations of radio
pulsars provide rough guidance about the time scales $\tau_{\mathrm
min}$.  In Paper~I we considered two fiducial classes of sources which
we denoted: (i) Young, fast pulsars, with $f_{\mathrm{max}}=1000$~Hz
and $\tau_{\mathrm{min}}=40$~yr, and (ii) old, slow pulsars, with
$f_{\mathrm{max}}=200$~Hz and $\tau_{\mathrm{min}}=1000$~yr.  To
facilitate direct comparison with the achievable sensitivities quoted
in Paper~I, we again use these two classes to present our results.

The two extremes of sky area to be searched are: (i) zero steradians for a
\emph{directed} search in which we know the source location in
advance; e.g.\ a supernova remnant, and (ii) $4\pi$~steradians for an
\emph{all-sky} search.  We consider both of these cases,
as well as the intermediate case of a 0.004~steradian
search about the galactic center.

Recent
work~\cite{Andersson_N:1998,Friedman_J:1997,Lindblom_L:1998,Owen_B:1998}
has suggested that new-born rapidly spinning neutron stars may evolve
on a time scale of months rather than decades, radiating away most of
their angular momentum in the form of gravitational waves within a
year.  These sources may be loud enough to be detected in other
galaxies, in which case optical detection of a supernova can serve as
a trigger for a targeted search.  Therefore we consider the case of a
directed search for sources with frequencies of
$f_{\mathrm{max}}$=200~Hz and evolution time scales of
$\tau_{\mathrm{min}}$=1~yr.

A final class of sources that we consider are accreting neutron stars
in binary systems.  Several such binary systems have been identified
via x-ray observations; the rotation frequencies of the accreting
neutron stars are inferred to be $\sim 250$--$350$~Hz
($f_{\mathrm{max}}$=700~Hz).  Bildsten~\cite{Bildsten_L:1998} has argued
that these accreting objects in low mass x-ray binaries (LMXB's) may
emit detectable amounts of gravitational radiation.  Since the
positions of these sources are well localized on the sky by their
x-ray emissions, the earth-motion induced Doppler modulations of the
gravitational waves can be precisely determined.  The difficulty with
these sources is the unknown, or poorly-known, orbits of the neutron
stars about their stellar companions, and the stochastic
accretion-induced variations in their spin.  We have estimated the
size of these effects, and outlined a search algorithm in
Sec.~\ref{ss:xray-binaries}.  These issues deserve further study in an
effort to improve the search strategy.

\subsection{Search Technique}
\label{ss:search-technique}

In searches for continuous gravitational waves, our sensitivity will
be limited by the computational resources available, rather than the
duration of the signal or the total amount of data.  Therefore the
computational efficiency of a search technique is extremely important.
For example, matched filtering (convolution of noise-whitened detector
output with a noise-whitened template) may detect a signal with the
greatest signal-to-noise ratio for any given stretch of data; however,
it becomes computationally prohibitive to search over large parameter
spaces with long data stretches.  A sub-optimal, but more efficient,
algorithm might achieve the best overall sensitivity for a fixed
amount of computational resources.  We present two possible search
strategies to accumulate signal to noise from the data stream.

Central to both of these methods is the technique we adopt to
demodulate the signal.  We can remove the effects of Doppler and
spindown modulations by defining a canonical time coordinate
\begin{equation}
\label{eq:tb}
t_b[t;\vec\lambda] = t + \frac{\vec x(t)}{c}\cdot\hat{n} +
	\sum_{k=1}\frac{f_k}{k+1}\left[t +
	\frac{\vec x(t)}{c}\cdot\hat{n}\right]^{k+1} \; ,
\end{equation}
with respect to which the signal, defined in Eq.~(\ref{eq:strain}), is
perfectly sinusoidal:
\begin{figure}[tb]
\vbox{
\psfig{file=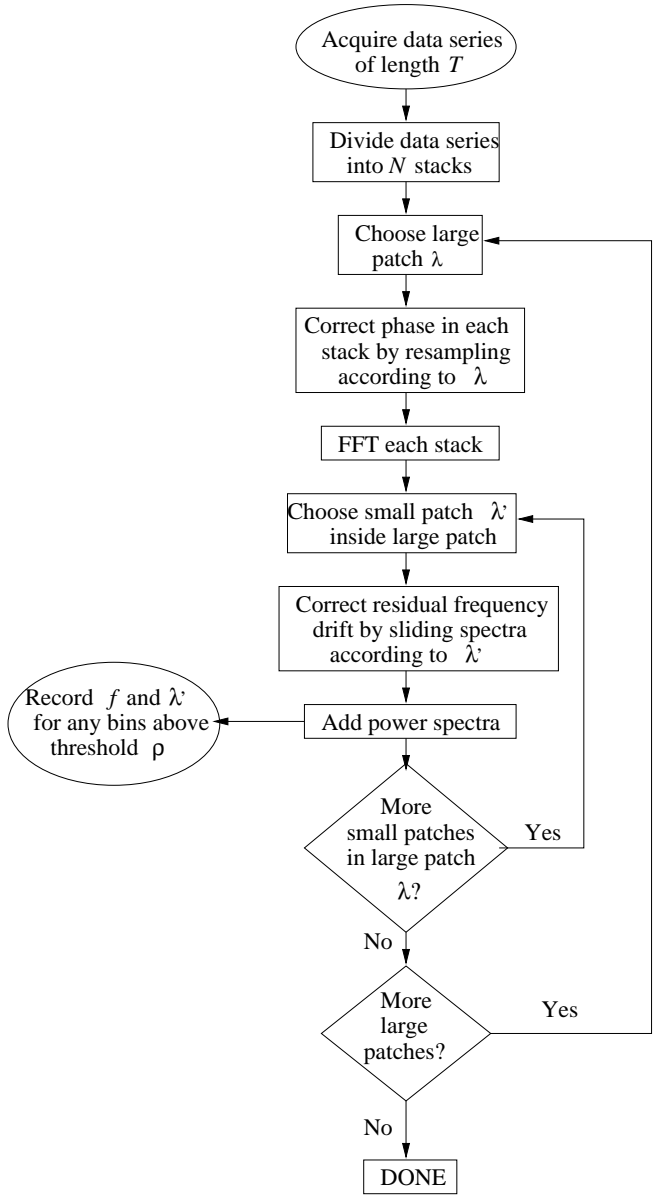,width=7.6cm,bbllx=150pt,%
bblly=70pt,bburx=455pt,bbury=740pt}
\caption{\label{fig:stacked-flowchart}
A flowchart representation of the {\em stacked slide} algorithm to
search for sources of continuous gravitational waves.   Notice that the
computational cost of sampling the fine grid is reduced by sliding the
power spectra,  rather than re-computing an FFT for each point on the
fine grid.
}}
\end{figure}\noindent
\noindent \begin{equation}
\label{eq:demodulated-strain}
h(t_b[t;\vec\lambda]) = {\cal A}e^{-2\pi i f_0 t_b[t;\vec\lambda]
	- i\Psi} \; .
\end{equation}
(Remember, we treat $\cal A$ and $\Psi$ as constant in time.)  The
introduction of the new time coordinate can be achieved by re-sampling
the data stream at equal intervals in $t_b$. The power spectrum,
computed from the Fourier transform of the re-sampled data, will
consist of a single monochromatic spike, whose amplitude (relative to
broadband noise) increases in proportion to the length of the data
stretch.  In practice the data will be sampled in the detector frame,
so that a sample may not occur at the desired value of $t_b$.
Consequently, we advocate the use of nearest-neighbor (stroboscopic)
resampling~\cite{Schutz_B:1991}.  This method will not substantially
reduce the signal to noise in a search provided the detector output is
sampled at a sufficiently high frequency. (See
Appendix~\ref{app:resampling-error}.)

When the waveform shape parameters $\vec\lambda$ are not known in
advance, one must search over a mesh of points in parameter space.
The result of a phase correction and Fourier transform will be
sufficiently monochromatic only if the true signal parameters lie
close enough to the one of the mesh points.  In Sec.~\ref{s:mismatch}
we rigorously define what is meant by ``close enough'', and show how
to determine the number of points for which corrections should be
applied.  We note that the approach of resampling followed
by a Fourier transform has the benefit that a single Fourier transform
automatically searches over all frequencies $f_0$, leaving only the
shape parameters $\vec\lambda$ to be searched explicitly.  Other
demodulation techniques, such as matched filtering, must apply
separate corrections for each value of $f_0$ in addition to the
$\vec\lambda$.  This increases the computational cost dramatically.

A signal can also be accumulated \emph{incoherently} from successive
stretches of data by adding their power
spectra~\cite{Anderson_S:1993}.  However, even if each data stretch is
demodulated to sufficient precision that the power from a signal is
focused in a single Fourier frequency bin, residual errors in
$\vec\lambda$ may cause the power to be at different frequencies
between successive spectra. 
A more precise knowledge of the phase evolution is required to correct
for this drift; i.e.\ a finer mesh in parameter space.  However, once
a set of parameter corrections $\Delta\vec\lambda$ is assumed, it is
relatively easy to correct for the frequency drift: successive power
spectra are shifted in frequency by a correction factor $\Delta f$,
where $\Delta f$ is computed by differencing
$f(t;\mbox{\boldmath${\lambda}$\unboldmath})$, in
Eq.~(\ref{eq:gw-frequency}), between the initial and corrected guesses
for $\vec\lambda$, as a function of the start time of each data
stretch. Once the spectra have been corrected by $\Delta f$, they can
be added together.  This accumulates signal-to-noise less efficiently
than coherent phase corrections and FFT's, but is computationally
cheaper.

\subsubsection{Stack-slide search} \label{sss:stacked-slide}
The search techniques that we consider in this paper are variants on
the following scheme.  First, the data 
stream is divided into shorter
lengths, called \emph{stacks}.  
Each stack is phase corrected and
FFT'ed, using a mesh of correction points sufficient to confine a
putative signal to $\sim1$ frequency bin in each stack.  The
individual power spectra are then corrected for residual frequency
drift using a finer parameter mesh suitable to remove phase
modulations over the entire data stretch.  The corrected power spectra
are summed, and searched for spikes which exceed some specified
significance threshold~\cite{Anderson_S:1993}.  The complete procedure
is summarized in the flowchart in Fig.~\ref{fig:stacked-flowchart}.

\begin{figure}[tb]
\vbox{
\psfig{file=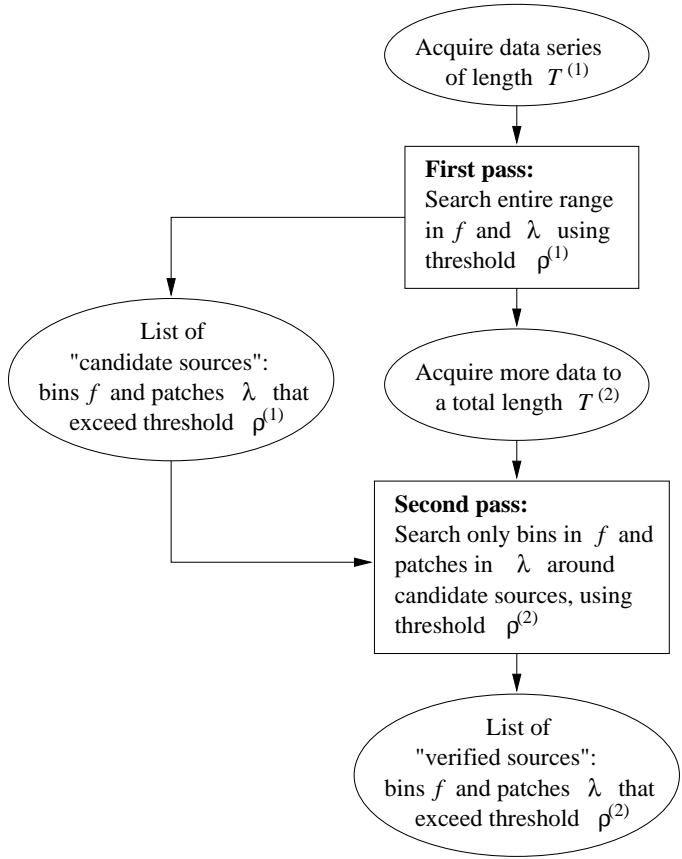,width=7.6cm,bbllx=174pt,%
bblly=201pt,bburx=428pt,bbury=580pt}
\caption{\label{fig:hierarchical-flowchart}
A flowchart representation of the {\em hierarchical} algorithm to
search for sources of continuous gravitational waves.   It should be
noted that while this approach will almost certainly be incorporated
into the eventual search algorithm for gravitational waves,  the real
benefit of such an approach will be to increase the confidence in a
detection made using some other technique.
}}
\end{figure}

\subsubsection{Hierarchical search} \label{sss:hierarchical}
We also consider a two-pass hierarchical search strategy.  In
this case, one performs an initial search of the data using a low
threshold which allows for many false alarms.  This is followed by
a second pass, using longer stretches of data, but searching the
parameter space only in the vicinity of the candidate detections of
the first pass.  This procedure is summarized in
Fig.~\ref{fig:hierarchical-flowchart}.  The advantage of a
hierarchical search are two-fold: (i) the low threshold on the first
pass allows detection of low-amplitude signals which would otherwise
be rejected, and (ii) the second pass can search longer data stretches
on a limited computing budget, because of the reduced parameter space
being searched, thus excluding false positives from the first pass.
For given computational resources,  this technique achieves the best
sensitivity of the strategies considered here and in Paper~I,  if
the thresholds and mesh points are optimally chosen between the
first and second passes.

\begin{figure}[tb]
\vbox{
\psfig{file=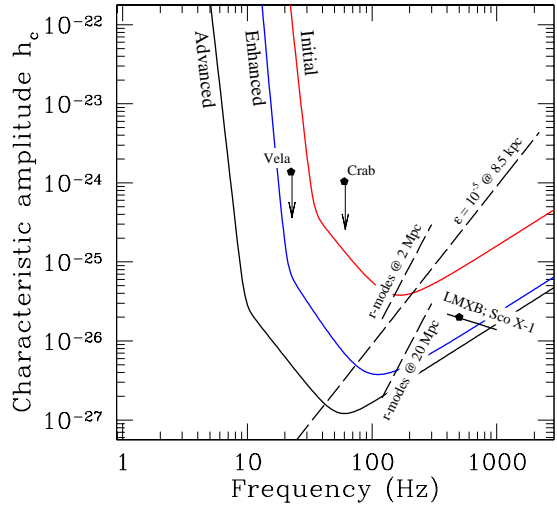,width=7.6cm,bbllx=18pt,%
bblly=164pt,bburx=592pt,bbury=718pt}
\caption{\label{fig:h_3peryear}
Characteristic amplitudes $h_c$ [see Eq.~(3.5)
in~\protect\cite{Brady_P:1998prd}] for several postulated periodic
sources, compared with sensitivities $h_{\rm \protect\scriptsize
3/yr}$ of the initial, enhanced and advanced detectors in LIGO.
($h_{\rm \protect\scriptsize 3/yr}$ corresponds to the amplitude $h_c$
of the weakest source detectable with 99\% confidence in $\frac{1}{3}
{\rm yr} = 10^7\mbox{\rm s}$ integration time, if the frequency and
phase of the signal, as measured at the detector, are known in
advance.)  Long-dashed lines show the expected signal strength as a
function of frequency for pulsars at a distance of 8.5~kpc assuming a
gravitational ellipticity $\epsilon = 10^{-5}$ of the source (see
Ref.~\protect\cite{Brady_P:1998prd}).  Upper limits are plotted
for the Crab and Vela pulsars, assuming their entire measured spindown
is due to gravitational wave emission.  The characteristic amplitude
of waves from $r$-modes is also shown.  These signals are not
precisely periodic; rather, they chirp downward through a frequency
band of $\sim200$~Hz in $2\times 10^7$ seconds.  Finally, the strength
of the gravitational waves from LMXB's, normalized to the observed
x-ray flux from Sco X-1, is plotted under the assumption that
gravitational waves are entirely responsible for their angular
momentum loss.  }}
\end{figure}\noindent

\subsection{Results}
\label{ss:results}

The sensitivity $\Theta = 1/h_{\mathrm{th}}$ of a search is defined in
Eq.~(\ref{eq:sensitivity}).  The threshold strain amplitude
$h_{\mathrm{th}}$ is defined such that there is a 1\% \textit{a
priori} probability that detector noise alone will produce an event
during the analysis, and therefore is the minimum characteristic
strain detectable
in the search.  We compare our results for the
sensitivity $\Theta$ to a canonical sensitivity determined by the
search threshold $h_{\mathrm{3/yr}}=4.2 \sqrt{S_n(f) \times 10^{-7}
\mathrm{Hz}}$.  This threshold is the characteristic amplitude of the
weakest source detectable with $99\%$ confidence in a coherent search
of $10^7$ seconds of data, if the frequency and phase evolution of the
signal are known.  The relative sensitivity $\Theta_{\mathrm{rel}}$ is
given by $\Theta_{\mathrm{rel}}\equiv
h_{\mathrm{3/yr}}/h_{\mathrm{th}}$; a relative sensitivity
$\Theta_{\mathrm{rel}}=0.1$ for a search means that a signal must have
a characteristic amplitude $h_c\agt10\times h_{\mathrm{3/yr}}$ to be
detected in that search.  Figure~\ref{fig:h_3peryear} shows
$h_{\mathrm{3/yr}}$ based on noise spectral estimates for three
detector systems in LIGO: the initial detectors are expected to go
on-line in the year 2000, with the first science run from 2002--2004;
the upgrade to the enhanced detectors should begin in $\sim$2004, with
subsequent upgrades leading to, and perhaps past, the advanced
detector sensitivity.  The expected amplitudes $h_c$ of several
putative sources are also shown; we use the definition of $h_c$ given
in Eq. (50) of Ref.~\cite{Thorne_K:1987}, and Eq.~(3.5) of Paper~I.
The strengths of gravitational waves from the Crab and Vela radio
pulsars are upper limits assuming all the rotational energy is lost
via gravitational waves.  The estimates of waves from the r-mode
instability are based on Owen \textit{et al.}~\cite{Owen_B:1998}, and
those from Sco X-1 are based on the recent analysis by
Bildsten~\cite{Bildsten_L:1998}.

\begin{figure}[tb]
\vbox{
\psfig{file=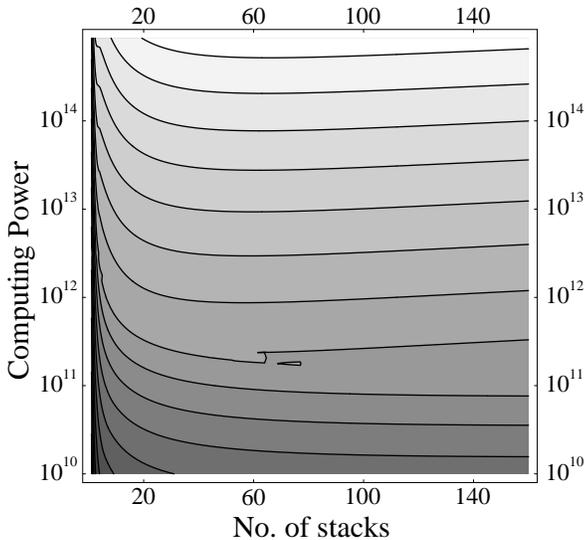,width=7.6cm,bbllx=62pt,%
bblly=158pt,bburx=540pt,bbury=618pt}
\caption{\label{fig:optimal}
A contour plot of the relative sensitivity defined in
Sec.~\protect\ref{ss:results} as a function of available
computing power, and the number of stacks in the search.  The plot
indicates the sensitivity that can be achieved using a stack-slide
search for sources with $f\leq 1000$ Hz and $\tau \geq 40$ yr.  The
darkest shading represents the worst sensitivity.  For fixed number of
stacks the sensitivity improves with increasing computing power as
expected.  Notice that for fixed computing resources there is
generally a point of optimal sensitivity; indeed the number
of stacks at this optimal operating point should be compared with
those given in Fig.~\protect\ref{fig:stacked-directed}.  It is
important to notice that the maximum falls off very slowly as the
number of stacks increases.}}
\end{figure}

A reasonable long-term search strategy is that data analysis should
proceed at roughly the same rate as data acquisition.  Given finite
computational resources and a desired overall false alarm probability,
there is an optimal choice for the length of a data stretch, and the
number of stacks that one should analyze in a given search run.  The
optimal strategy is that which \emph{maximizes} the final
sensitivity of a search subject to the constraints on computational
resources and time to analyze the data.  We have
plotted the relative sensitivity of a search for young, fast pulsars
as a function of the number of stacks and the available computing
power in Fig.~\ref{fig:optimal}.  The optimal number of stacks is
easily read off the plot for fixed computing power.  Note that
the maximum sensitivity in this plot is quite flat, especially in the
regime where one is most computationally bound.  This may be extremely
relevant when implementing these search techniques; data management
issues may impose more severe constraints on the size and number of
stacks than computational power does.  This remains to be explored
when the data analysis platforms have been chosen.

\begin{figure}[tb]
\vbox{
\vbox{
\vskip -0.6in
\psfig{file=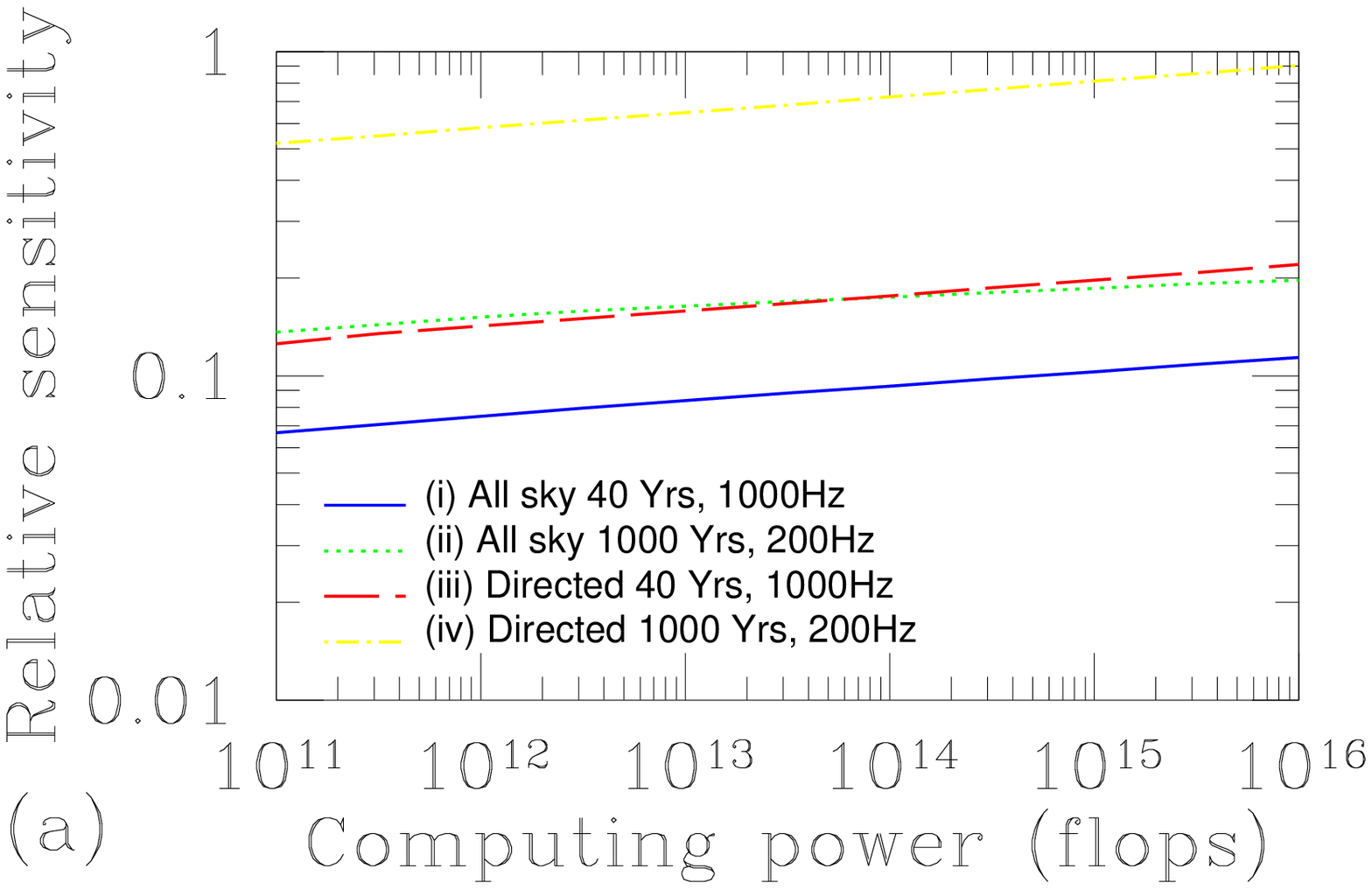,width=7.6cm,bbllx=18pt,%
bblly=244pt,bburx=472pt,bbury=700pt}
\vskip -1.1in
\psfig{file=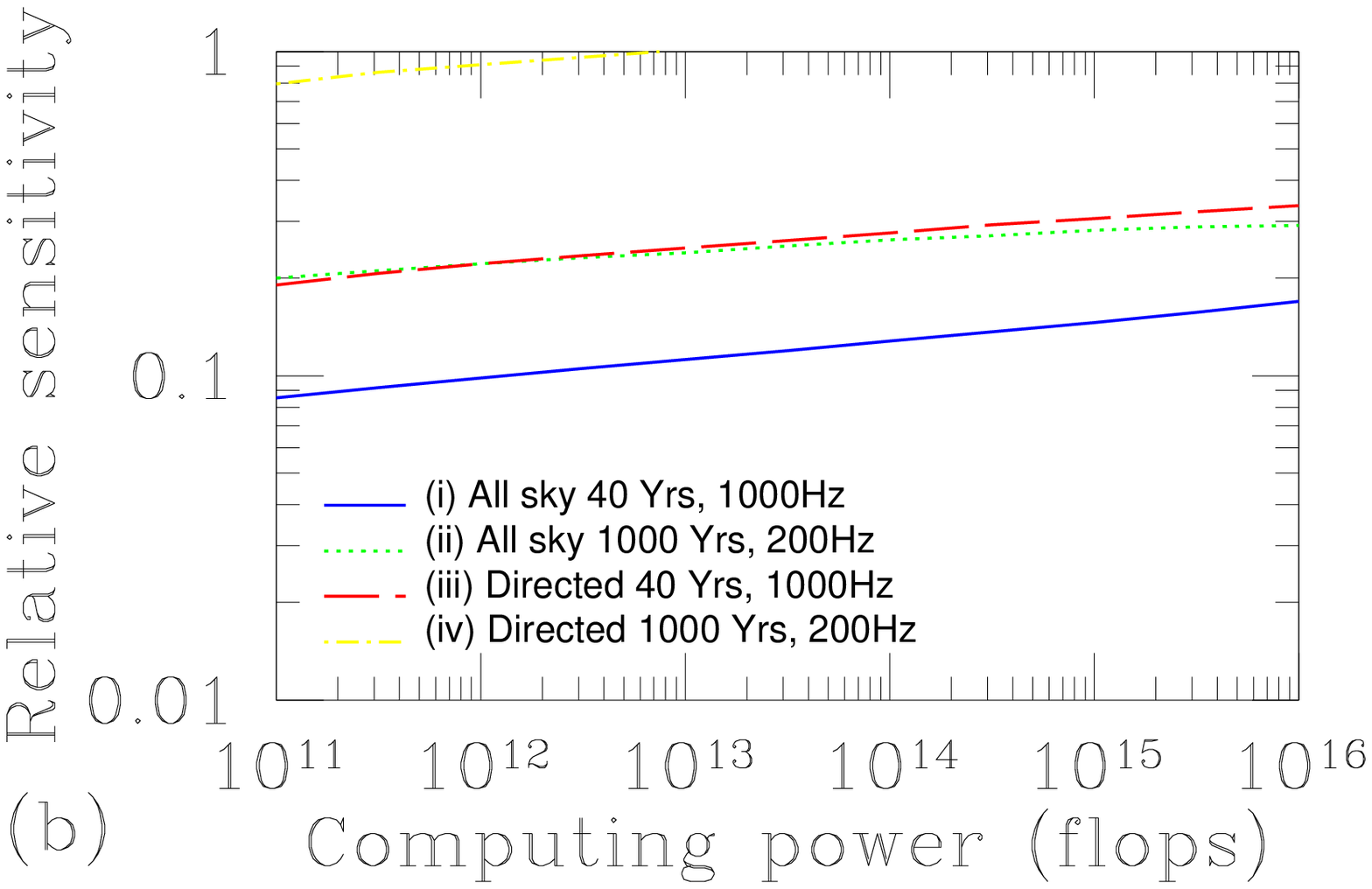,width=7.6cm,bbllx=18pt,%
bblly=250pt,bburx=472pt,bbury=718pt}
\vskip-0.3in}
\caption{\label{fig:stacked-sensitivities} Relative amplitude
sensitivities
$\Theta_{\mathrm{rel}}=h_{\mathrm{3/yr}}/h_{\mathrm{th}}$ achievable
with given computational resources, for (a) one pass stack-slide
search strategies, and (b) two-pass hierarchical strategies (using the
stack-slide algorithm in each pass).  The results are presented for
our fiducial classes of sources: (i)~all-sky search for young
($\tau\ge 40$~yr), fast ($f\le 1000$~Hz) pulsars, (ii)~all-sky search
for old($\tau\ge 1000$~yr), slow ($f\le 200$~Hz) pulsars,
(iii)~directed search for young, fast pulsars, and, (iv)~directed
search for old, slow pulsars.  For a given computational power, we
have determined the optimum observation time, and number of stacks as
described in Secs.~\protect\ref{s:stack-search} and
\protect\ref{s:hierarchical-search-with-stacking}.  Thus
$h_{\mathrm{th}}$ is the expected sensitivity of the detector for an
optimal
stack-slide search, with 99\% confidence.}}
\end{figure}

Figure~\ref{fig:stacked-sensitivities}(a) shows the optimal
sensitivities which can be achieved, as a function of available
computing power, using a stack-slide search.  The results are
presented for both fiducial classes of pulsars: old ($\tau\geq
1\,000$~yr) slow ($f \leq 200$~Hz) pulsars, and young ($\tau\geq
40$~yr) fast ($f\leq 1\,000$~Hz) pulsars.  In each case, we have
considered both directed and all-sky searches for the sources.  The
results should be compared with those of Paper~I, in which we
considered coherent searches without stacking: the use of stacked
searches gains a factor of $\sim$2--4 in sensitivity.

The use of a two-pass hierarchical search can further improve
sensitivities by balancing the computational requirements between the
two passes.  Figure~\ref{fig:stacked-sensitivities}(b) shows
the sensitivities achievable when each pass uses a stack-slide
strategy.  The sensitivities achieved exceed those of one-pass
stack-slide searches by $\sim 20$--$60$\%.

\begin{figure}[tb]
\vbox{
\vbox{
\psfig{file=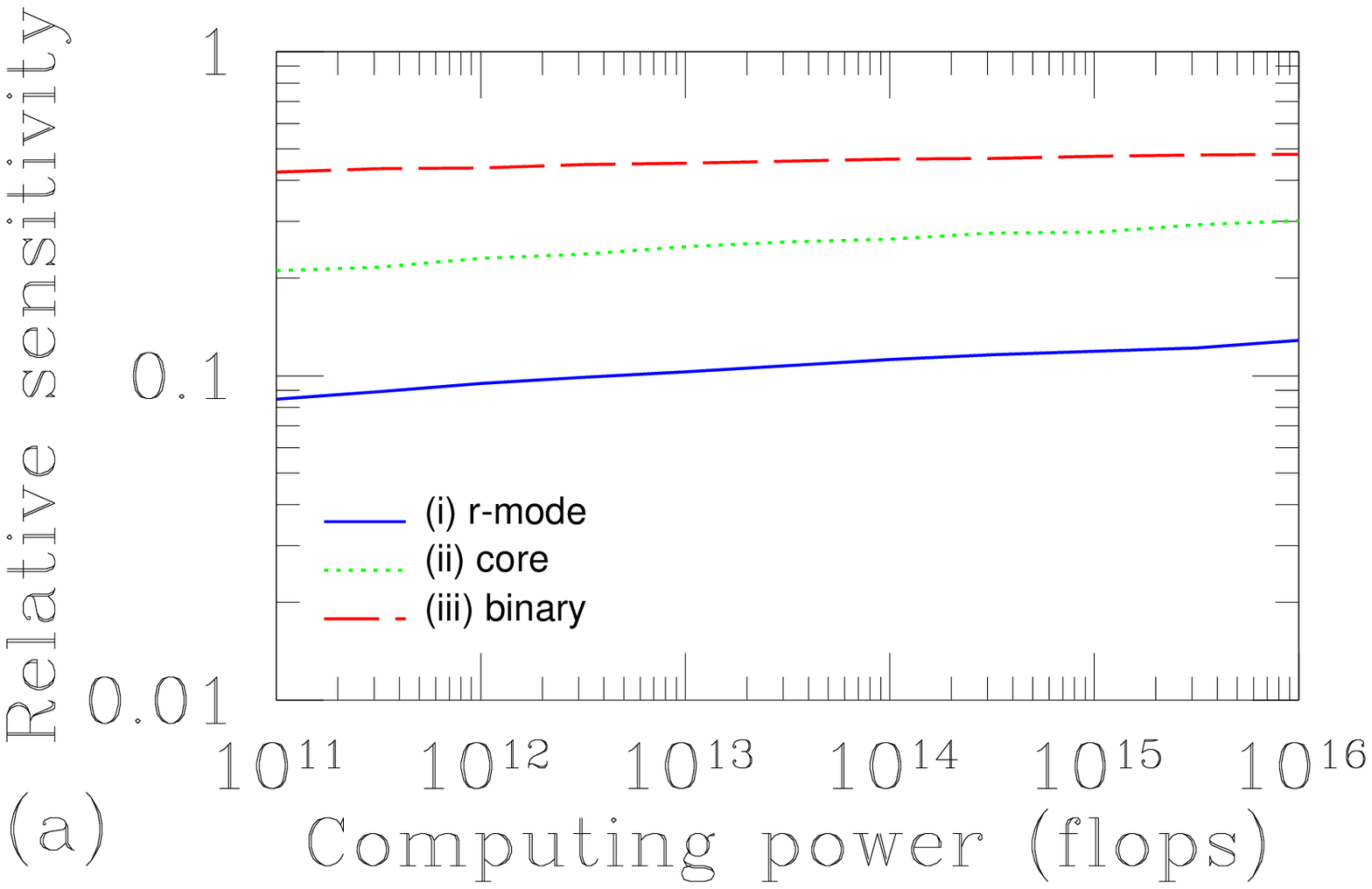,width=7.6cm,bbllx=18pt,%
bblly=144pt,bburx=472pt,bbury=610pt}
\vskip -1.8in
\psfig{file=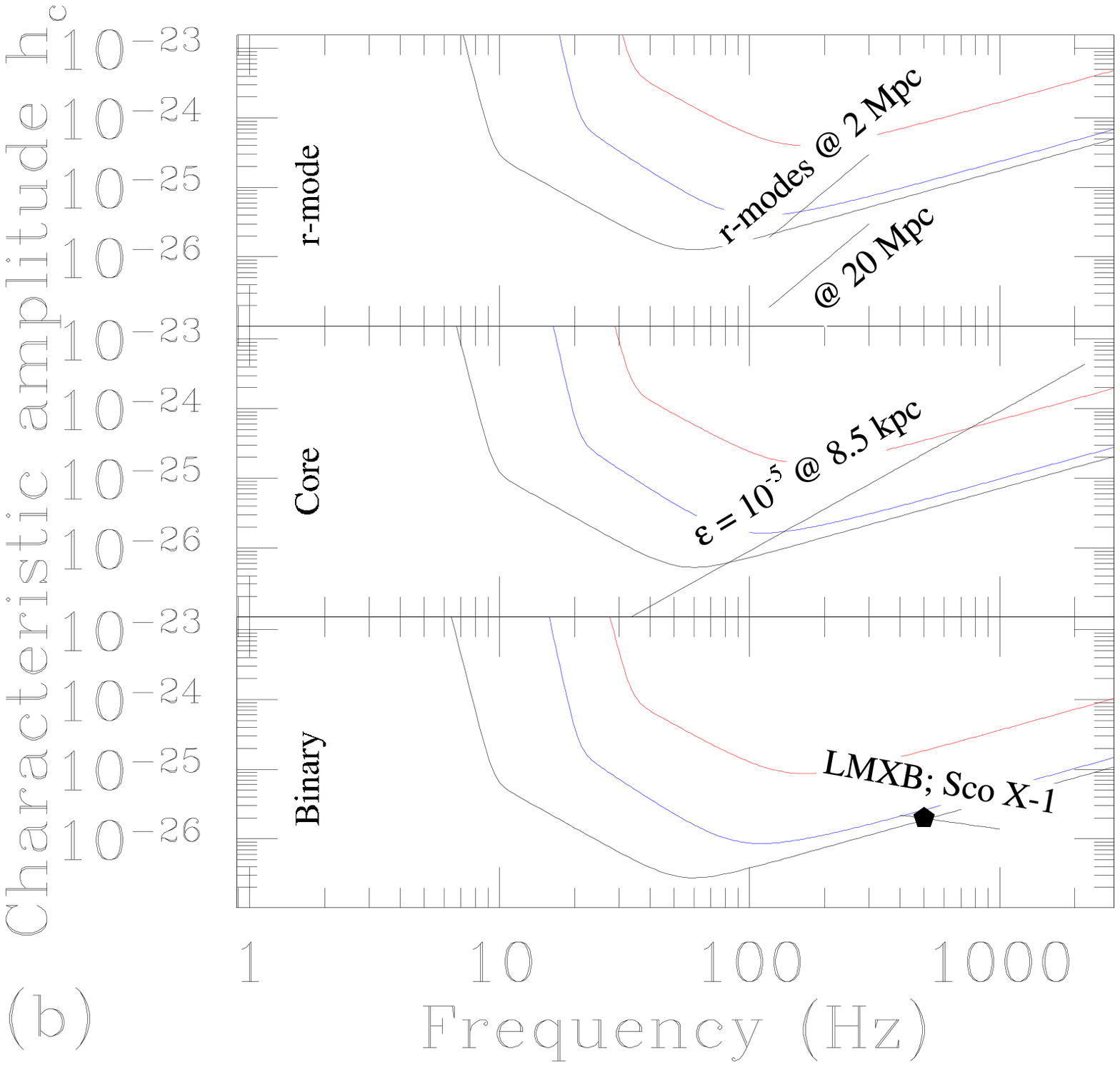,width=7.6cm,bbllx=18pt,%
bblly=144pt,bburx=472pt,bbury=718pt}
\vskip -0.1in
}
\caption{\label{fig:targeted-sensitivities} Panel (a) represents the
relative amplitude sensitivities
$\Theta_{\mathrm{rel}}=h_{\mathrm{3/yr}}/h_{\mathrm{th}}$ achievable with
given
computational resources, in three specialized searches: (i) A search
for a new-born neutron star (whose direction is determined by
observing an optical supernova) that is spinning down by gravitational
wave emission via an active $r$-mode instability.  We took $\tau\ge
1$~yr and $f\le 200$~Hz.  (ii) A search for pulsars in a region
extending $0.004$ steradians about the galactic core, with $\tau\ge
100$~yr and $f\le 500$~Hz.  (iii) A source in a binary orbit, e.g.
Sco~X-1.  We assume the orbit is characterized by two orthogonal
velocity parameters, known to within a total error of
$17(\mathrm{km/s})^2$; we further assume that the frequency
$f\leq500$~Hz experiences a random walk typical of Eddington-rate
accretion.  For each of these sources, panel (b) shows
$h_{\mathrm{th}}$ for initial (upper lines), enhanced (middle lines),
and advanced (lower lines) interferometers in LIGO, assuming
$1$~Tflops of computing power.  Thus this is the characteristic
amplitude of the weakest source that can be detected with $99\%$
confidence using a two pass-hierarchical search strategy.  }}
\vskip -0.1in
\end{figure}

The computational requirements for all-sky, all frequency surveys are
sufficiently daunting that we explore three restricted searches in
Sec.~\ref{s:specialized-searches}: (i) a directed search for a newborn
neutron star in the young ($\lesssim1$~year old) remnant of an
extra-galactic supernova, (ii) an area search of the galactic core for
pulsars with $\tau \geq 100\> \textrm{yr}$ and $f \leq
500\>\textrm{Hz}$, and (iii) a directed search for an accreting
neutron star in a binary system (such as Sco X-1).
Figure~\ref{fig:targeted-sensitivities} shows the relative
sensitivities attainable in such searches.  With computational
resources capable of $1$ Tflops, we expect to see galactic core
pulsars with enhanced LIGO if they have non-axisymmetric strains of
$\epsilon\agt 5\times10^{-6}$ at frequencies of $\sim200$~Hz.
Estimates of the characteristic strain of gravitational waves from an
active $r$-mode instability in a newborn neutron star suggest that
these sources will be detectable by the enhanced interferometers in
LIGO out to distances $\sim 8$~Mpc; the rate of supernovae is $\sim
0.6$ per year within this distance.  Finally, gravitational waves from
accreting neutron stars in low-mass x-ray binary systems (LMXBs) may
be detectable by enhanced interferometers in LIGO if we can obtain
sufficient information about the binary orbit from electromagnetic
observations.  Sco~X-1 is on the margins of detectability using the
enhanced LIGO interferometers operating in broadband configuration.
We estimate that the amplitude signal-to-noise from these sources
could be improved by a factor of $\sim 5$--$10$ by operating the
interferometer in a signal-recycled, narrow-band configuration.

\subsection{Organization of the paper}
\label{ss:organization}

In Sec.~\ref{s:mismatch} we extend the metric formalism that was
developed in Paper~I to determine the number of parameter space points
that must be sampled in a search that accumulates signal to noise by
summing up power spectra.   This method can then be used to compute
the number of correction points needed in a stack-slide search.
Approximate formulae,  useful for estimating the computational cost of a
search,  are presented for the number of corrections needed in an
all-sky search,  and also in directed searches of a single sky
position.

We discuss the the issue of thresholding in
Sec.~\ref{s:thresholds-and-sensitivities}.  Then we present the
computational cost estimates, and determine the optimal parameters for
single-pass, stack-slide searches in Sec.~\ref{s:stack-search}

Section~\ref{s:hierarchical-search-general-remarks} presents a general
discussion of hierarchical searches for periodic sources using a
single interferometer.  Schutz~\cite{Schutz_B:GWDAW} has previously
considered hierarchical searches demonstrating their potential in
searches for periodic sources.  The relationship between the threshold
in the second stage of the search, and the threshold required in the
first stage is discussed in detail.  We also present the computational
cost of each stage of the search.  These results are used in
Sec.~\ref{s:hierarchical-search-with-stacking} to determine the
optimal search parameters in hierarchical searches for our fiducial
classes of sources.

Finally, we discuss three specialized searches in
Sec.~\ref{s:specialized-searches}.  We present a preliminary
investigation of issues which arise when the gravitational wave source
is in a binary system (e.g.\ an LMXB).  We discuss a parameterized
model of the binary orbit, and estimate the number of parameter space
points which must be sampled in a search for the gravitational waves
from one of the objects in the binary.  In the case where the emitter
is accreting material from its companion, we also allow for stochastic
changes in frequency due to fluctuations in the accretion rate.

Detailed formulae for the number of points in parameter space when
dealing with a stacked search are presented in
Appendix~\ref{s:patch-number-formulae}.  In
Appendix~\ref{app:resampling-error} we discuss the loss in signal to
noise that can occur when using nearest neighbor resampling to apply
corrections to the detector output.  If the data is sampled at
$16\,384$~Hz, we demonstrate that this method will lose less than
$1\%$ of amplitude signal-to-noise for a signal with gravitational
wave frequency $\leq 1000$~Hz,

\section{Mismatch}
\label{s:mismatch}

In a detection strategy that searches over a discrete mesh of points
in parameter space, the search parameters and signal parameters will
never be precisely matched.  This mismatch will reduce the signal to
noise since the signal will not be precisely monochromatic.  It is
desirable to quantify this loss, and to choose the grid spacing so
that the loss is within acceptable limits.  This can be achieved by
defining a distance measure on the parameter space based on the
fractional losses in detected signal power due to parameter mismatch.
In Paper~I we derived such a measure in the case where the search was
performed using coherent Fourier transforms; this method was modeled
after Owen's computation of a metric on the parameter space of
coalescing binary waveforms~\cite{Owen_B:1996}.  In this paper we
extend this approach to the case of incoherent searches, in which
several power spectra are added incoherently, or {\it stacked}, and
then searched for spikes.

Let $h(t;\mbox{\boldmath${\lambda}$\unboldmath})$ be a hypothetical
signal given by Eq.~(\ref{eq:strain}) with true signal parameters
$\mbox{\boldmath${\lambda}$\unboldmath}=(f_0,\vec\lambda)$.  If the
data containing this signal are corrected for some nearby set of shape
parameters $\vec\lambda+\Delta\vec\lambda$, the signal will take the
form
\begin{equation}
\label{eq:mismatched-strain}
h_b(t;\mbox{\boldmath${\lambda}$\unboldmath},
\Delta\mbox{\boldmath${\lambda}$\unboldmath}) =
{\cal A} e^{-i\{2\pi f_0 t_b[t;\vec\lambda]
	+ \phi[t;\mbox{\scriptsize\boldmath${\lambda}$\unboldmath}]
	- \phi[t;(f_0,\vec\lambda+\Delta\vec\lambda)]\}} \; ,
\end{equation}
where the subscript $b$ is used to indicate the corrected waveform.
In a stacked search, the data are divided into $N$ segments of equal
length $\Delta T$, each of these segments is Fourier transformed, and
then a total power spectrum
$P_h(f;\mbox{\boldmath${\lambda}$\unboldmath},\Delta\vec\lambda)$ is
computed according to the formula
\begin{equation}
\label{eq:stacked-power}
H(f;\mbox{\boldmath${\lambda}$\unboldmath},\Delta\vec\lambda) = 2
\sum_{k=1}^N
	|\tilde
h_k(f;\mbox{\boldmath${\lambda}$\unboldmath},\Delta\vec\lambda)|^2 \; .
\end{equation}
The Fourier transform of each individual segment is defined to be
\begin{equation}
\label{eq:h-tilde}
\tilde h_k(f;\mbox{\boldmath${\lambda}$\unboldmath},\Delta\vec\lambda)
= \frac{\cal A}{\sqrt{\Delta T}}\int_{(k-1)\Delta T}^{k\Delta T}
	e^{i\Delta\phi[t;\mbox{\scriptsize\boldmath${\lambda}$\unboldmath},
\Delta\mbox{\scriptsize\boldmath${\lambda}$\unboldmath}]}  \; dt_b
\end{equation}
where $\Delta\phi$ is given by
\begin{equation}
\label{eq:delta-phase}
\Delta\phi[t;\mbox{\boldmath${\lambda}$\unboldmath},
\Delta\mbox{\boldmath${\lambda}$\unboldmath}] = 2\pi (f-f_0)t_b
	+ \phi[t;(f_0,\vec\lambda+\Delta\vec\lambda)]
	- \phi[t;\mbox{\boldmath${\lambda}$\unboldmath}] \; .
\end{equation}
Here $\Delta\mbox{\boldmath${\lambda}$\unboldmath} =
(f-f_0,\Delta\vec\lambda)$ denotes the error in matching the
modulation shape parameters \emph{and} the error in sampling the
resulting power spectrum at the wrong frequency.  Both of these errors
lead to a reduction in the detected power relative to the optimum case
where the carrier frequency and the phase modulation are precisely
matched.

The \emph{mismatch} $m(\mbox{\boldmath${\lambda}$\unboldmath},
\Delta\mbox{\boldmath${\lambda}$\unboldmath})$, which is the
fractional reduction in power due to imperfect phase correction
\emph{and} sampling at the wrong Fourier carrier frequency, is defined
to be
\begin{equation}
\label{eq:mismatch-def}
m(\mbox{\boldmath${\lambda}$\unboldmath},
\Delta\mbox{\boldmath${\lambda}$\unboldmath}) = 1 -
\frac{H(f;\mbox{\boldmath${\lambda}$\unboldmath},\Delta\vec\lambda)}
	{H(f_0;\mbox{\boldmath${\lambda}$\unboldmath},\vec 0)} \; .
\end{equation}
Remember $\mbox{\boldmath${\lambda}$\unboldmath} = (\lambda_0,
\vec{\lambda}) = (f, \lambda_1, \lambda_2 \ldots)$.
Substituting the expressions for $H$ from Eq.~(\ref{eq:stacked-power})
into Eq.~(\ref{eq:mismatch-def}), we find
\begin{equation}
\label{eq:mismatch-stacked}
m(\mbox{\boldmath${\lambda}$\unboldmath},
\Delta\mbox{\boldmath${\lambda}$\unboldmath}) = 1 - 
\frac{1}{N{\cal A}^2}\sum_{k=1}^N |\tilde 
h_k(f;\mbox{\boldmath${\lambda}$\unboldmath},\Delta\vec\lambda)|^2 \; .
\end{equation}
It is easily shown that $m(\mbox{\boldmath${\lambda}$\unboldmath}, 
\Delta\mbox{\boldmath${\lambda}$\unboldmath})$ has a local minimum of
zero when $\Delta\mbox{\boldmath${\lambda}$\unboldmath} = 0$.  We
therefore expand the mismatch in powers of
$\Delta\mbox{\boldmath${\lambda}$\unboldmath}$ to find
\begin{equation}
\label{eq:mismatch-metric}
m(\mbox{\boldmath${\lambda}$\unboldmath},
\Delta\mbox{\boldmath${\lambda}$\unboldmath}) = 
\sum_{\alpha,\beta} g_{\alpha\beta}(\mbox{\boldmath${\lambda}$\unboldmath})
	\Delta\lambda^\alpha \Delta\lambda^\beta
	+ {\cal O}(\Delta\mbox{\boldmath${\lambda}$\unboldmath}^3) \; ,
\end{equation}
where $(\alpha,\beta)$ are summed over $0,1,\ldots,j,\ldots$.
The quantity $g_{\alpha\beta}$ is a local distance {metric} on
the parameter space.  This metric is explicitly given by
\begin{equation}
\label{eq:metric-def}
g_{\alpha\beta}(\mbox{\boldmath${\lambda}$\unboldmath}) \equiv
{\textstyle\frac{1}{2}}
\partial_{\Delta\lambda^\alpha}\partial_{\Delta\lambda^\beta}
m(\mbox{\boldmath${\lambda}$\unboldmath},
\Delta\mbox{\boldmath${\lambda}$\unboldmath})
|_{\Delta\mbox{\scriptsize\boldmath${\lambda}$\unboldmath} = 0} \; ,
\end{equation}
where $\partial_{\Delta\lambda^\alpha}$ denotes a partial derivative
with respect to $\Delta\lambda^\alpha$.
It is convenient to express $g_{\alpha\beta}$ as a sum of metrics
computed for the individual stacks,  that is
\begin{equation}
\label{eq:metric-stacked-total}
g_{\alpha\beta}(\mbox{\boldmath${\lambda}$\unboldmath}) =
\frac{1}{N}\sum_{k=1}^N
	g_{\alpha\beta}^{(k)}(\mbox{\boldmath${\lambda}$\unboldmath})
\end{equation}
where the individual stack metrics
$g_{\alpha\beta}^{(k)}(\mbox{\boldmath${\lambda}$\unboldmath})$
are explicitly given by
\begin{equation}
\label{eq:metric-stacked-individual}
g_{\alpha\beta}^{(k)}(\mbox{\boldmath${\lambda}$\unboldmath}) =
	\langle\partial_{\Delta\lambda^\alpha}\Delta\phi \,
		\partial_{\Delta\lambda^\beta}\Delta\phi\rangle_k -
	\langle\partial_{\Delta\lambda^\alpha}\Delta\phi\rangle_k
	\langle\partial_{\Delta\lambda^\beta}\Delta\phi\rangle_k \; .
\end{equation}
The phase error $\Delta\phi$ is given in Eq.~(\ref{eq:delta-phase}),
and we use the notation
\begin{equation}
\label{eq:average-def}
\langle\ldots\rangle_k = \left.\frac{1}{\Delta T}
	\int_{(k-1)\Delta T}^{k\Delta T} (\ldots) dt_b
	\right|_{\Delta\mbox{\scriptsize\boldmath${\lambda}$\unboldmath}=0} \; .
\end{equation}

In a search, we will look for spikes in the power spectrum
$H(f;\mbox{\boldmath${\lambda}$\unboldmath},\Delta\vec{\lambda})$ computed
from the detector output,
that is, we will look for local maxima in the frequency parameter $f$.
Therefore the relevant measure of distance in the space of shape
parameters $\vec{\lambda}$ is the fractional loss in power due to
mismatched parameters $\Delta\vec\lambda$, but after maximizing over
frequency.  We therefore define the \emph{projected mismatch}
$\mu(\vec\lambda,\Delta\vec\lambda)$ to be
\begin{equation}
\label{eq:projected-mismatch}
\mu(\vec\lambda,\Delta\vec\lambda) =
	\min_{f}
m(\mbox{\boldmath${\lambda}$\unboldmath},
\Delta\mbox{\boldmath${\lambda}$\unboldmath}) = \sum_{i,j}
	\gamma_{ij}(\vec\lambda)\Delta\lambda^i\Delta\lambda^j \; ,
\end{equation}
where
\begin{equation}
\label{eq:projected-metric}
\gamma_{ij} = \left( g_{ij} - \frac{g_{i0}g_{j0}}{g_{00}}
	\right)_{\lambda^0=f_{\mathrm{max}}}
\end{equation}
is the mismatch metric projected onto the subspace of shape
parameters, and $f_{\mathrm{max}}$ is the maximum frequency that we
include
in the search.   The meaning of the minimization $\min_f$ is clear
from the definition of the mismatch in Eq.~(\ref{eq:mismatch-stacked}).

The distance function, and in particular the metric in
Eq.~(\ref{eq:projected-metric}), can be used to determine the number
of discrete mesh points that must be sampled in a search.  Let $\cal
P$ be the space of all parameter values $\vec\lambda$ to be searched
over, and define the maximal mismatch $\mu_{\mathrm{max}}$ to be the
largest
fractional loss of power that we are willing to tolerate from a
putative source with parameters in $\cal P$.  For the model waveform
in Eqs.~(\ref{eq:gw-frequency}), (\ref{eq:gw-phase}), and
(\ref{eq:strain})  this parameter space is coordinatized by
$\vec{\lambda} = (\theta,\phi,f_1,f_2,\ldots)$ where $\theta$,$\phi$
denote location of the source on the sky,  and $f_j$ are related to
the time derivative of the intrinsic frequency of the source.  Each
correction
point of the mesh is considered to be at the center of a cube with
side $2\sqrt{\mu_{\mathrm{max}}/n}$, where $n$ is the dimension of $\cal
P$;
this insures that all points in $\cal P$ are within a proper distance
$\mu_{\mathrm{max}}$ of a discrete mesh point as measured with the metric
$\gamma_{ij}$.  The number of patches required to fill the parameter
space is
\begin{equation}
\label{eq:n-patches}
N_p(\Delta T,\mu_{\mathrm{max}},N) = \frac{\int_{\cal P}
	\sqrt{\det |\gamma_{ij}|}d^n\lambda}
	{(2\sqrt{\mu_{\mathrm{max}}/n})^n} \; .
\end{equation}
Since $\mu_{\mathrm{max}}$ is the maximum loss in detected
power \emph{after} the power spectra have been added,   $N_p(\Delta
T,\mu_{\mathrm{max}},N)$ is the number of patches required to construct
the
\emph{fine} mesh in the stacked search strategy described in
Sec.~\ref{ss:search-technique}.  The coarse mesh in the stacked search
strategy requires only that the spikes in the individual power spectra
be reduced by no more than $\mu_{\mathrm{max}}$; consequently, the number
of
points in such a mesh is simply $N_p(\Delta T,\mu_{\mathrm{max}},1)$.

We note that the average expected power loss for a source randomly
placed within such cubical patches is $\langle\mu\rangle=
\mu_{\mathrm{max}}/3$.  (In Paper I we quoted an average that was
computed for ellipsoidal patches;  this is not appropriate to the
cubical grid which will likely be used in a real search.)

\subsection{Directed search}
\label{ss:directed-search}

In most cases, the forms of Eqs.~(\ref{eq:gw-phase}) and (\ref{eq:tb})
are sufficiently complicated to defy analytical solution, especially
since $\vec v$ in Eq.~(\ref{eq:gw-phase}) should properly be taken
from the true ephemeris of the Earth during the period of observation.
However, for the case of a directed search, that is a search in just a
single sky direction, the phase correction is polynomial in $t$, and
the metric can be computed analytically.  To a good approximation, the
metric is flat --- the spacing of points in parameter space is
independent of the value of the spindown parameters
$(f_1,f_2,\ldots,f_s)$. For a given number $s$ of spindown parameters
in a search, the right hand side of Eq.~(\ref{eq:n-patches}) can be
evaluated analytically.  The result is expressed as a product ${\cal
N}_s {G}_s$, where
\begin{equation}
\label{eq:ns-directed}
{\cal N}_s = \frac{f_{\mathrm{max}}^s (\Delta T)^{s(s+3)/2}}
	{(\mu_{\mathrm{max}}/s)^{s/2}\tau_{\mathrm{min}}^{s(s+1)/2}} \;
\end{equation}
depends on the maximum frequency $f_{\mathrm{max}}$ (in Hz), the
length of each stack $\Delta T$ (in seconds), the maximal mismatch
$\mu_{\mathrm{max}}$, and the minimum spindown age
$\tau_{\mathrm{min}}$ (in seconds) to be considered in the search.
The dependence on the number of stacks $N$ is contained in $G_s(N)$
which are given by:
\begin{eqnarray}
\label{eq:g0}
G_0(N) &=& 1 \; ,\\
\label{eq:g1}
G_1(N) &\approx& 0.524N \; ,\\
\label{eq:g2}
G_2(N) &\approx& 0.0708N^3 \; ,\\
\label{eq:g3}
G_3(N) &\approx& 0.00243N^6 \; ,
\end{eqnarray}
when $N \gg 4$.   The detailed expressions for $G_s(N)$ are presented
in Appendix~\ref{s:patch-number-formulae}.  For up to $3$ spindown terms
in the
search,  the number of patches is then
\begin{equation}
\label{eq:np-directed}
N_p(\Delta T,\mu_{\mathrm{max}},N)
 = \max_{s\in\{0,1,2,3\}} [{\cal N}_s G_s(N)] \; .
\end{equation}
The maximization accounts for the situation where increasing $s$,  the
dimension of the parameter space,  decreases the value of ${\cal N}_s
G_s$ because the parameter space extends less than one patch width in
the new spindown coordinate $f_s$;   one should not search over this
coordinate.

\subsection{Sky search}
\label{ss:sky-search}

For signal modulations that are more complicated than simple power-law
frequency drift, it is impossible to compute $N_p$ analytically.  In
an actual search over sky positions as well as spindown, one should
properly compute the mismatch metric numerically, using the exact
ephemeris of the Earth in computing the detector position.  In Paper~I
we computed $N_p(T_b,\mu_{\mathrm{max}},1)$ numerically, with the
simplification that both the Earth's rotation and orbital motion were
taken to be circular.  However, in this paper we are concerned also
with the dependence of $N_p$ on the number of stacks $N$.  This
significantly complicates the calculation of the metric and its
determinant,  and makes it necessary to adopt some approximations in
the calculation.  Fortunately, the results of interest here are rather
insensitive to errors in $N_p$.

In Paper~I we mentioned that there are strong correlations between sky
position and spindown parameters, thus requiring the use of the full
$s+2$~dimensional metric.  However, these correlations are due
primarily to the Earth's orbital motion, to which a low-order Taylor
approximation is good for times much less than a year.  Therefore we
treat the number of patches as the product of the number of spindown
patches times the number of sky positions ${\cal M}_s$, computed
analytically using only the Earth's rotational motion.  We note that
this approximation is appropriate only for computing the number of
patches; \emph{when actually demodulating the signals, the true
orbital motion would have to be included}.  This approximation works
well so long as the orbital \emph{residuals} (the remaining orbital
modulations after correction on this sky mesh) are much smaller than
the spindown corrections being made at the same power in $t$.  The
residual orbital velocity at any power $t^k$ is roughly
\begin{equation}
\label{eq:residual}
\sim \frac{\alpha_k}{k!\sqrt{{\cal M}_s}}\times\frac{r\Omega}{c}
	\times(\Omega t)^k \; ,
\end{equation}
where $\alpha_k$ is a number of order unity, ${\cal M}_s$ is the
number of sky patches, and $r=1$AU and $\Omega=2\pi$/yr are the
Earth's orbital radius and angular velocity.  When the range in this
residual is comparable to or larger than the range in the
corresponding spindown term $f_kt^k$, the ``spindown'' parameter space
must be expanded to include the orbital residuals.  While the range in
$\alpha_k$ is difficult to arrive at analytically, we have found that
assuming a maximum value of $\approx0.3$ gives good agreement with the
numerical results of Paper~I (i.e.\ for $N=1$), to within factors of
$\sim2$.

One other approximation was made in computing the number of sky
patches.  We found that the measure $\sqrt{\det|\gamma_{ij}|}$ for the
sky position metric is almost constant in the azimuth $\varphi$, and
has a polar angle dependence which is dominantly of the form
$\sin2\theta$.  When performing the integral over sky positions, we
approximated the measure by $\sqrt{\det |\gamma_{ij}|} \simeq
\textrm{constant} \times \sin 2\theta$;  this approximation is accurate
to about one part in $10^4$.

Given these approximations, the number of patches for a sky search is
given by:
\begin{equation}
\label{eq:np-sky}
N_p = \max_{s\in\{0,1,2,3\}} \left[{\cal M}_s \overline{{\cal N}}_s G_s
	\prod_{k=0}^s\left(1+\frac{0.3r\Omega^{k+1}\tau_{\mathrm{min}}^k}
	{c\,k!\sqrt{{\cal M}_s}}\right)\right] \; .
\end{equation}
The number of sky patches ${\cal M}_s$, in the
$(s+2)$-dimensional search, is given approximately by
\begin{eqnarray}
\label{eq:ms-sky}
{\cal M}_s &\approx& \frac{f_{\mathrm{max}}^2}{4\mu_{\mathrm{max}}/(s+2)}
	\left(A^{-2}+B^{-2}+C^{-2}\right)^{-1/2} \; , \\
\label{eq:a-sky}
A &=& 0.014 \; , \\
\label{eq:b-sky}
B &=& 0.046 (\Delta T/1\,\mathrm{day})^2 \; , \\
\label{eq:c-sky}
C &=& 0.18 (\Delta T/1\,\mathrm{day})^5 N^3 \; .
\end{eqnarray}
This is a fit to the analytic result given in
Appendix~\ref{s:patch-number-formulae}.  The number of spindown
patches $\overline{{\cal N}}_s G_s$ in the $(s+2)$-dimensional search
is
\begin{equation}
\overline{{\cal N}}_s G_s = \frac{s^{s/2}}{(s+2)^{s/2}}
{{\cal N}}_s G_s \;
\end{equation}
where ${{\cal N}}_s$ and $G_s$ are given in
Eqs.~(\ref{eq:ns-directed})--(\ref{eq:g3}), and the prefactor on the
right corrects for the sky dimensions.  The remaining product terms
$\prod$ in Eq.~(\ref{eq:np-sky}) represent the increase in the size of
the spindown space in order to include the orbital residuals.

\section{Thresholds and sensitivities}
\label{s:thresholds-and-sensitivities}

The thresholds for a search are determined under the assumption
that the detector noise is a stationary, Gaussian random process with
zero mean and power spectral density $S_n(f)$.  In the absence of a
signal, the power $P_n(f)=2|\tilde{n}(f)|^2$ at each sampled frequency
is exponentially distributed with probability density function
$e^{-P_n/S_n}/S_n$.  The statistic for stacked spectra
is $\rho = \sum_1^N P_n(f)$.   The cumulative probability
distribution function for $\rho$,  in the absence of a signal, is
\begin{equation}
\label{eq:probability}
\textrm{CDF}[\rho/S_n,N] = \int_0^{\rho/S_n} e^{-r}
	\frac{r^{N-1}}{(N-1)!} \; dr
	= \frac{\gamma(N,\rho/S_n)}{(N-1)!}
\end{equation}
where $\gamma(N,\rho/S_n)$ is an incomplete gamma function.

A (candidate) detection occurs whenever $\rho$ in some frequency
bin exceeds a pre-specified threshold $\rho_c$ chosen so that the
probability of a false trigger due to noise alone is small.  There are
$f_{\mathrm{max}}\Delta T$ Fourier bins in each spectrum, and $N_p(\Delta
T,
\mu_{\mathrm{max}},N)$
spectra in the entire search.  Therefore we assume that a search
consists of $N_p f_{\mathrm{max}}\Delta T$ independent trials of the
statistic $\rho$,  and compute the expected number of false events $F$
to be
\begin{equation}
\label{eq:false-events}
F = f_{\mathrm{max}}\Delta T \, N_p(\Delta T,\mu_{\mathrm{max}},N)\,
	(1 - \textrm{CDF}[\rho_c/S_n,N])  \; .
\end{equation}
(In reality, there will be correlations between the statistic computed
for different frequencies and different patches.  Since this will
reduce the number of independent trials, Eq.~(\ref{eq:false-events})
overestimates the number of false events.  This is a small effect
which should not change the overall sensitivity of a search by much.
It is only in the case that the number of trials is initially small
that one should be concerned with this effect; unfortunately, we
operate in the other extreme.)  If $F \ll 1$,  number of false events
$F$ is approximately
equal to the probability that an event is caused by noise in the
detector.  Consequently, $\alpha=1-F$ can be thought of as
our confidence of detection.  In a non-hierarchical search, the
threshold $\rho_c$ is set by specifying $\alpha$ and then inverting
Eq.~(\ref{eq:false-events}).

Finally, how does the threshold $\rho_c$ affect the
sensitivity of our search?  We define a threshold amplitude
$h_{\mathrm{th}}$ to be the minimum dimensionless signal amplitude that we
expect to register as a detection in the search
\begin{equation}
\label{eq:strain-threshold}
h_{\mathrm{th}} = \sqrt{\frac{(\rho_c/N - S_n)}
	{\langle F_+^2(\Theta,\Phi,\Psi)\rangle
	(1-\langle\mu\rangle)\Delta T}}
\end{equation}
where $\langle F_+^2(\Theta,\Phi,\Psi)\rangle$ is the square of the
detector response averaged over all possible source positions and
orientations, and
$\langle\mu\rangle=\mu_{\mathrm{max}}/3$ is the expected mismatch of a
signal which is randomly located within a patch.  The
\emph{sensitivity} $\Theta$ of the search is then defined by
\begin{equation}
\label{eq:sensitivity}
\Theta \equiv \frac{1}{h_{\mathrm{th}}} \propto
	\sqrt{\frac{(1-\mu_{\mathrm{max}}/3)\Delta T}
	{\rho_c/N - S_n}} \; .
\end{equation}
For any optimal search strategy, the goal of optimization will be to
maximize the final sensitivity of the search, given limited
computational power.

\section{Stack-slide search}
\label{s:stack-search}

A stack-slide search is the simplest alternative to coherent searches
we consider here.  The main steps involved in the algorithm are shown
in the flow chart of Fig.~\ref{fig:stacked-flowchart}.  In this
section we estimate the computational cost of each step, and determine
the ultimate sensitivity of this technique.

The first step, before the search begins, is to specify the size of
the parameter space to be searched (i.e.\ choose $f_{\mathrm{max}}$,
$\tau$,
and a region of the sky), the computational power $P$ that will be
available to do the data analysis, and an acceptable false alarm
probability.  From these, one can determine optimal values for the
maximum mismatch $\mu_{\mathrm{max}}$ for a patch, the number of stacks
$N$, and the length $\Delta T$ of each stack, using the optimization
scheme discussed at the end of this section.  For now, we treat these
as free parameters.

Coarse and fine grids are laid down on the parameter space with
$N_{pc}=N_p(\Delta T,\mu_{\mathrm{max}},1)$ and $N_{pf}=N_p(\Delta
T,\mu_{\mathrm{max}},N)$ points, respectively.  The data-stream is
low-pass
filtered
to the upper cutoff frequency $f_{\mathrm{max}}$, and broken into $N$
stretches of length $\Delta T$.  Each of the steps above have
negligible computational cost since they are done only once for the
entire search.  The subsequent steps, on the other hand, must be executed
for each of the $N_{pc}$ correction points.

Each stretch of data is re-sampled (at the Nyquist frequency
$2f_{\mathrm{max}}$) and simultaneously demodulated by stroboscopic
sampling
for a set of demodulation parameters selected from the coarse grid.
The result is $N$ demodulated time series, each one consisting of
$n=2f_{\mathrm{max}}\Delta T$ samples.  Since stroboscopic demodulation
only
shifts one in every few thousand data points (assuming a sampling rate
at the detector of $16\,384\>\mathrm{Hz}$), the computational cost of
the demodulation itself is negligible.

Each stretch of data is then Fourier transformed using a fast Fourier
transform (FFT) algorithm with a computational cost of
$3nN\log_2(n)$~floating point operations.  Power spectra are computed
for each Fourier series, costing 3 floating point operations per
frequency bin, i.e., a total cost of $1.5nN$~floating point
operations.

For demodulation parameters in the coarse grid, the power of a matched
signal will be confined to $\sim1$ Fourier bin in each power spectrum,
but not necessarily the \emph{same} bin in different spectra.  To
insure that power from a signal is accumulated by summing the $N$
spectra, we must apply $N_{pf}/N_{pc}$ corrections within each coarse
mesh patch.  This can be achieved by the following steps.

For the $N$ spectra to be stacked, the frequency of a putative signal
with initial frequency $f_{\mathrm{max}}$ is computed using
Eq.~(\ref{eq:gw-frequency}).  Each spectrum is re-indexed so that the
power from such a signal would be in the same frequency bin (we ignore
the computational cost of this step), and the spectra are added [$0.5
n (N-1)$ floating point operations].  We automatically account for
corrections at other frequencies by applying the fine grid corrections
in this way.  It may be possible to reduce the computational cost of
this portion of the search by noting, for example, that we over count
the fine grid corrections for signals with frequency
$f_{\mathrm{max}}/2$ by a factor of $2^n$ where $n$ is the dimension
of the parameter space being explored.  Since it is difficult to
assess the feasibility of using this in a real search,   we simply
mention it so that it might be explored at the time of
implementation.

The resulting stacked spectrum is scanned for peaks which exceed
the threshold $\rho_c$.  Since this has negligible computational cost,
the number of floating point operations required for the entire search
is
\begin{equation}
\label{eq:computational-cost}
C = 3nNN_{pc}\, [\log_2(n) + 0.5 + N_{pf}(N-1)/(6NN_{pc})] \; .
\end{equation}

If data analysis proceeds at the same rate as data
acquisition, the computational power $P$ required to complete a search
is $P=C/N\Delta T$ floating-point operations 
\begin{figure}[tb]
\vbox{
\vbox{
\psfig{file=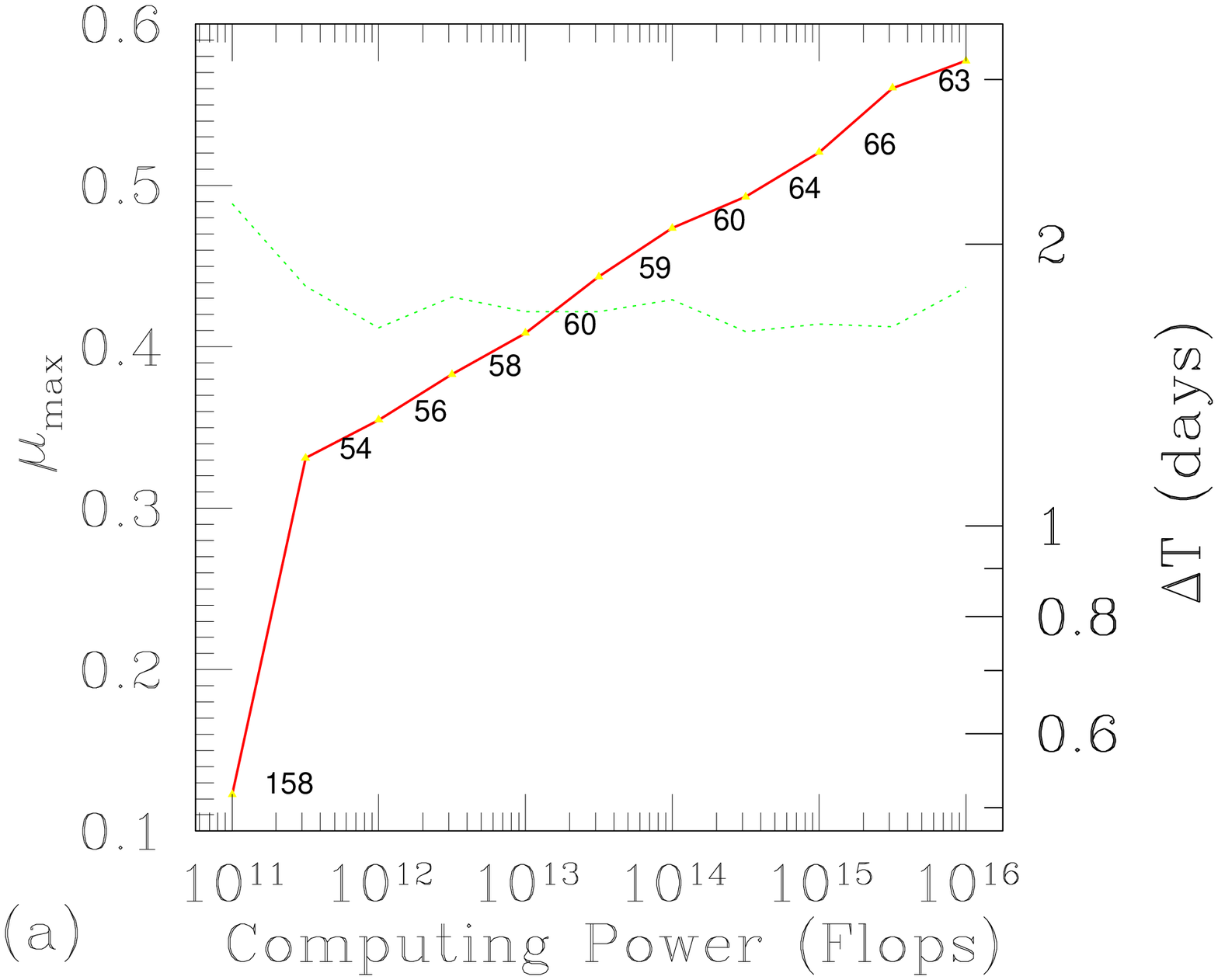,width=9cm,bbllx=0in,%
bblly=1.8in,bburx=8.5in,bbury=8.5in}\vskip -0.2in
\psfig{file=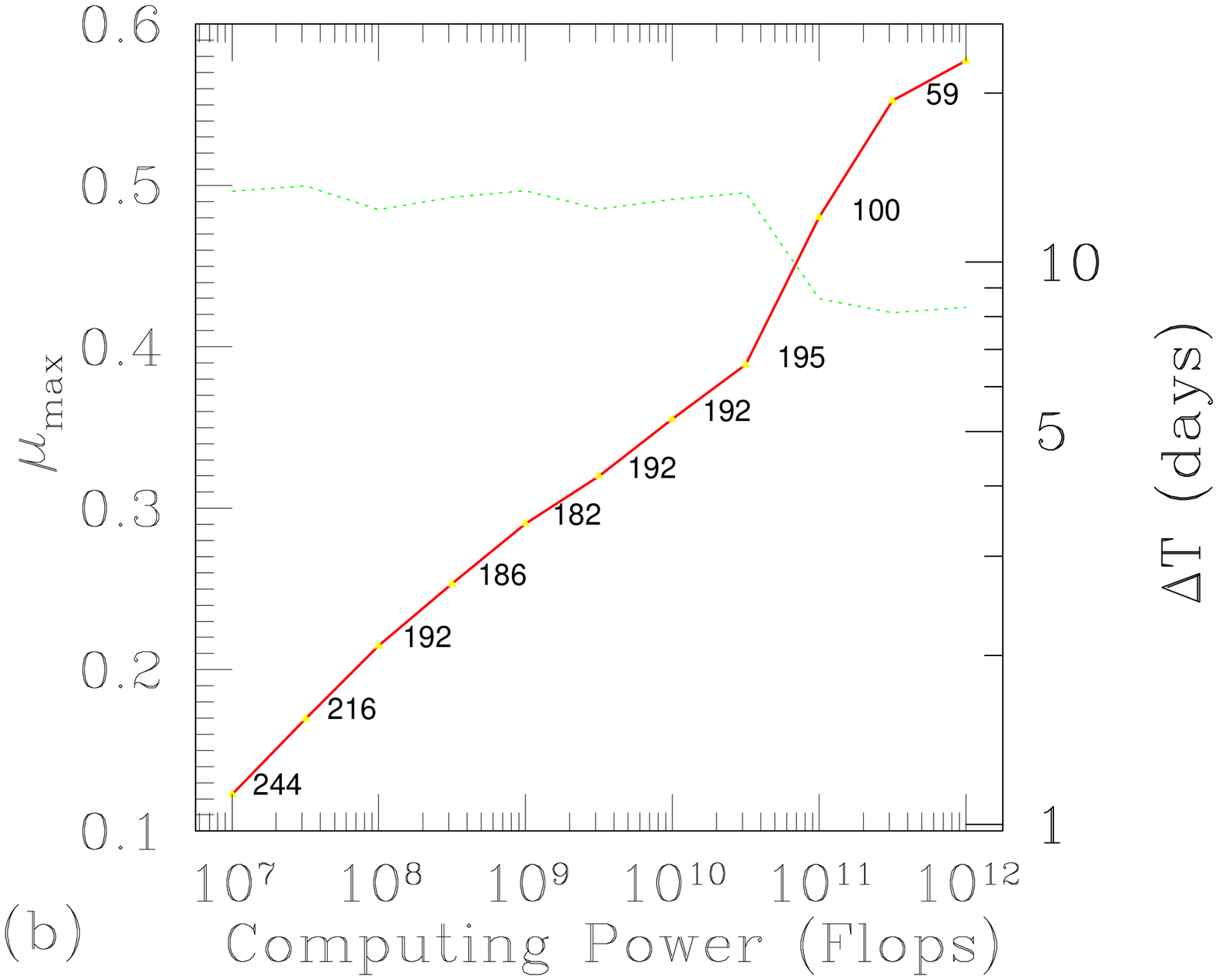,width=9cm,bbllx=0in,%
bblly=1.8in,bburx=8.5in,bbury=8.5in}
}
\caption{\label{fig:stacked-directed} The optimum stack length
$\Delta T$ (thick solid line), number of stacks $N$ (numbers along the
solid line), and maximal projected mismatch $\mu_{\mathrm{max}}$ (dotted
line) as functions of available computational power for directed,
stack-slide searches.  The two panels correspond to the following
cases: (a) The situation encountered when searching for periodic
sources having gravitational wave frequencies up to $1000\mbox{\rm
Hz}$, with minimum spindown ages $\tau_{\rm \protect\scriptsize
min}=40\mbox{\rm yr}$.  (b) The equivalent results for gravitational
wave frequencies up to $200\mbox{\rm Hz}$, with minimum spindown ages
$\tau_{\rm \protect\scriptsize min}=10^3\mbox{\rm yr}$.  Both cases
assume a threshold which gives an overall statistical significance of
$99\%$ to a detection (although the results are insensitive to the
precise value).  The optimization was performed numerically using
simulated annealing which accounts for some of the fluctuations in the
observation times.}}
\end{figure}\noindent
per second (Flops).
Equation~(\ref{eq:computational-cost}) and the definition of $n=2
f_{\mathrm{max}}\Delta T$ imply that the computational power is
\begin{equation}
\label{eq:computational-power}
P = 6f_{\mathrm{max}} N_{pc}\, [\log_2(n) + 0.5 + N_{pf}(N-1)/(6NN_{pc})]
\; .
\end{equation}

The final sensitivity $\Theta$, defined in Eq.~(\ref{eq:sensitivity}),
of the search is determined once we know the function $N_p$, the
\begin{figure}[tb]
\vbox{
\vbox{
\psfig{file=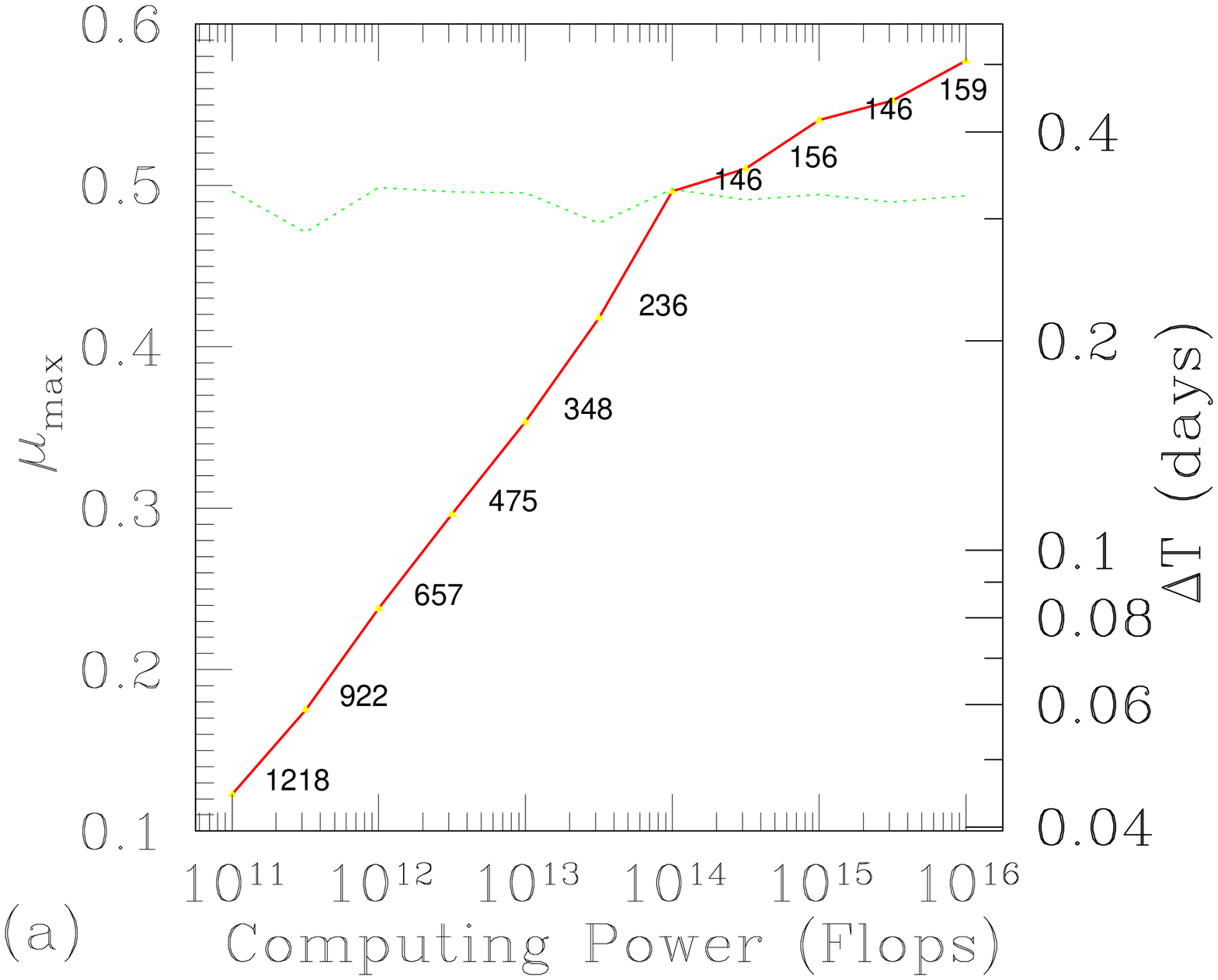,width=9cm,bbllx=0in,%
bblly=1.8in,bburx=8.5in,bbury=8.5in}\vskip -0.2in
\psfig{file=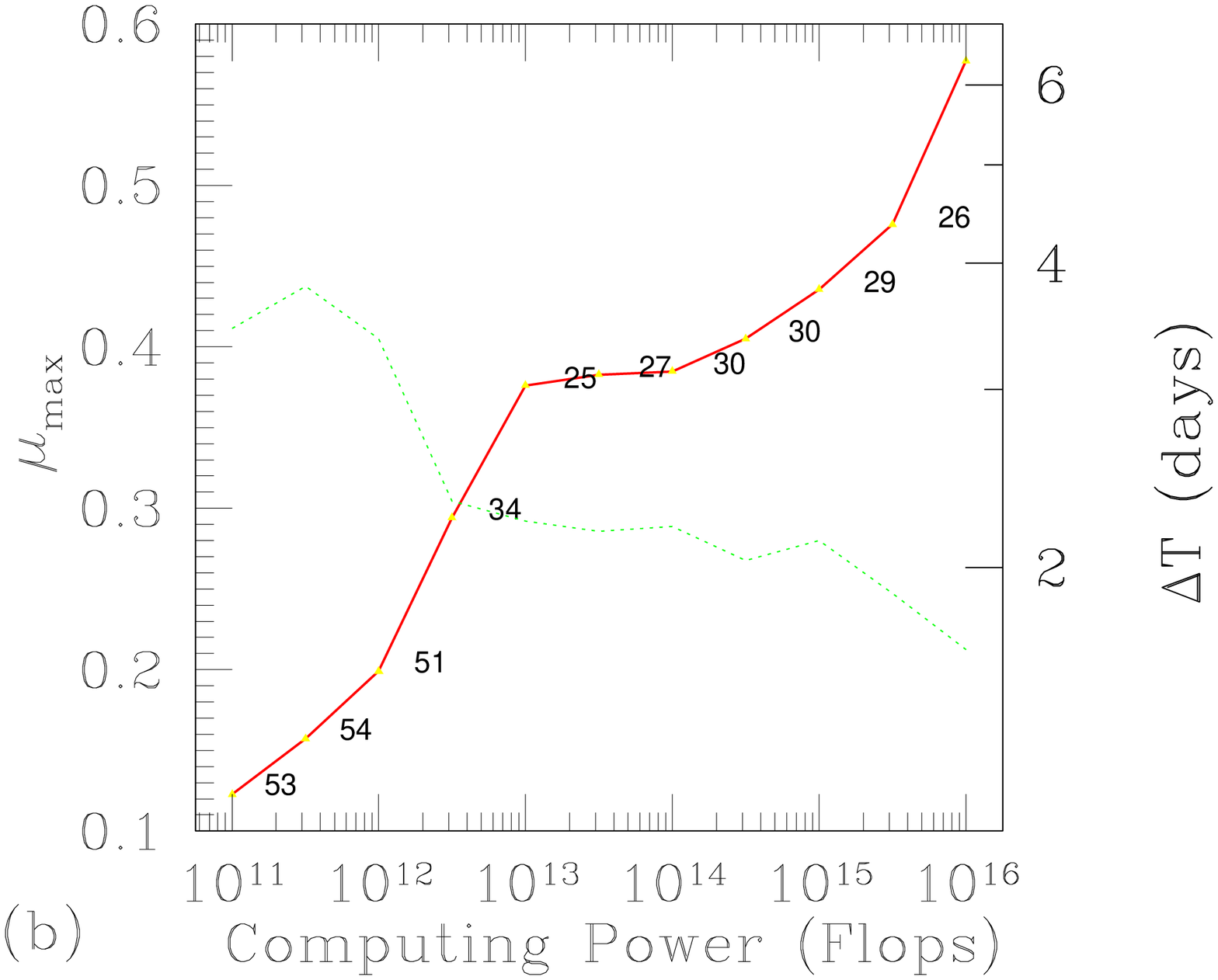,width=9cm,bbllx=0in,%
bblly=1.8in,bburx=8.5in,bbury=8.5in}
}
\caption{\label{fig:stacked-all-sky}
Same as in Fig.~\ref{fig:stacked-directed}, but for an all-sky search.
}}
\end{figure}\noindent
frequency $f_{\mathrm{max}}$, the maximal mismatch $\mu_{\mathrm{max}}$,
and the
confidence level $\alpha=1-F$.  An optimized algorithm will maximize
$\Theta$ as a function of $\mu_{\mathrm{max}}$, $N$, and $\Delta T$,
subject to the constraints imposed by fixing the false alarm
probability $F$, and the computational power $P$.

The results of the optimization procedure are given in
Figures~\ref{fig:stacked-directed} and~\ref{fig:stacked-all-sky} for
the fiducial classes of pulsar defined in
Sec.~\ref{ss:parameter-ranges}.  In each case we have set the
probability of a false alarm threshold at $F=0.01$ (indicating a 99\%
confidence that detector noise will not produce an event above
threshold), and have determined optimal values of $\mu_{\mathrm{max}}$,
$N$,
and $\Delta T$ for a range of values of the available computational
power.  Figures~\ref{fig:stacked-directed}(a) and (b) show the results
for a directed search for old, slow ($\tau_{\mathrm{min}}$=1\,000~yr,
$f_{\mathrm{max}}$=200~Hz) and young, fast ($\tau_{\mathrm{min}}$=40~yr,
$f_{\mathrm{max}}$=1\,000~Hz) pulsars, respectively.
Figures~\ref{fig:stacked-all-sky}(a) and (b) show the results for an
all-sky search for the same two classes of source.  The optimal
sensitivities achieved by these searches are summarized in
Fig.~\ref{fig:stacked-sensitivities}(a) of the Introduction.

\section{Hierarchical search: general remarks}
\label{s:hierarchical-search-general-remarks}

The basic hierarchical strategy involving a two pass search is
represented schematically in Fig.~\ref{fig:hierarchical-flowchart}.
In the first pass, $N^{(1)}$ stacks of data of
length $\Delta T^{(1)}$ are demodulated on a coarse and fine mesh of
correction points computed for some mismatch level $\mu^{(1)}$, and
then searched by stacked Fourier transforms.  A threshold
signal-to-noise level is chosen which will, in general, admit many
false alarms.  In the second stage, $N^{(2)}$ stacks of length $\Delta
T^{(2)}$ are searched on a finer mesh of points computed at a mismatch
level $\mu^{(2)}$, but only in the vicinity of those events which
passed the first-stage threshold.  The second stage will involve fewer
correction points than the first, so the second-stage transforms can
be made longer and more sensitive.  The goal of optimization is
to find some combination of $\Delta T^{(1)}$, $\Delta T^{(2)}$,
$\mu^{(1)}$, $\mu^{(2)}$, $N^{(1)}$, and $N^{(2)}$ which maximizes the
final sensitivity for fixed computational power $P$,  and second
pass false alarm probability $F^{(2)}$.

\subsection{Thresholds}
\label{ss:thresholds}

In the first pass of a hierarchical search,  each of
$N_f^{(1)}=f_{\mathrm{max}}\Delta T^{(1)}$ frequency bins in
$N_{pf}^{(1)}=N_p(\Delta T^{(1)},\mu^{(1)},N^{(1)})$ stacked power
spectra will be scanned for threshold crossing events.
If (as we assume) all of these trials are statistically independent, the
number of false events above the threshold $\rho^{(1)}$ will be
\begin{equation}
\label{eq:hierarchical-false-triggers}
F^{(1)} =
N_{pf}^{(1)}N_f^{(1)}(1-\textrm{CDF}[\rho^{(1)}/S_n^{(1)},N^{(1)}])
	\; .
\end{equation}
We assume that the number of false events will significantly exceed the
number of true signals in this pass,   consequently the number of
events to be analyzed in the second pass will be $F^{(1)}$.

The second stage uses a coarse grid with $N_{pc}^{(2)}=N_p(\Delta
T^{(2)},\mu^{(2)},1)$ points,  and a fine grid with
$N_{pf}^{(2)}=N_p(\Delta
T^{(2)},\mu^{(2)},N^{(2)})$ points. On average each false alarm will
require $N_{pc}^{(2)}/N_{pf}^{(1)}$ coarse grid points, and
$N_{pf}^{(2)}/N_{pf}^{(1)}$ fine grid points in the second stage.
(When a second-pass mesh is coarser than the first pass's parameter
determination, the corresponding ratio should be taken as unity.)
Furthermore, since the first stage will identify the candidate
signal's frequency to within $\sim2$ frequency bins, the second-stage
search should be over the $2\Delta T^{(2)}/\Delta T^{(1)}$
second-stage frequency bins which lie in this frequency range.  Once
again, we assume the noise in all frequency bins (and over all grid
points) is
independent, so the number of false events which exceed
the threshold $\rho^{(2)}$ in the second stage is
\begin{eqnarray}
\label{eq:hierarchical-false-alarm}
1-\alpha &=& F^{(2)} \nonumber \\
        &=& 2F^{(1)}\frac{N_{pf}^{(2)}}{N_{pf}^{(1)}}
	\frac{\Delta T^{(2)}}{\Delta T^{(1)}}
	(1-\textrm{CDF}[\rho^{(2)}/S_n^{(2)},N^{(2)}]) \nonumber \\
	&=& 2 f_{\mathrm{max}} \Delta T^{(2)} N_{pf}^{(2)}
	(1-\textrm{CDF}[\rho^{(1)}/S_n^{(1)},N^{(1)}]) \nonumber \\
	&& \times (1-\textrm{CDF}[\rho^{(2)}/S_n^{(2)},N^{(2)}])\; ,
\end{eqnarray}
where $\alpha$ is our desired confidence level for the overall search.

The thresholds $\rho^{(1)}$ and $\rho^{(2)}$ cannot be assigned
independently; rather, they should be chosen so that any \emph{true}
signal buried in the noise that would exceed (in expectation value)
the second-stage threshold will have passed the first-stage
threshold.  In other words, it serves no purpose to set $\rho^{(2)}$
any lower than the weakest signal which would have passed
$\rho^{(1)}$.  A signal which is expected to pass the
second-stage threshold exactly has an amplitude
$|\tilde{h}^{(2)}|^2=\rho^{(2)}-N^{(2)}S_n^{(2)}$.
We define the \emph{false dismissal probability} $D$ to be
the probability that such a signal will be falsely
rejected in the first pass.  Since the spectral power of a true signal
increases with $N\Delta T$, the signal seen in the first pass has
amplitude $|\tilde{h}^{(1)}|^2 = |\tilde{h}^{(2)}|^2 (N^{(1)}\Delta
T^{(1)})/(N^{(2)}\Delta T^{(2)})$,  and the thresholds satisfy the
relation
\begin{eqnarray}
\label{eq:hierarchical-false-dismissal}
D & = & \textrm{CDF}\left[\frac{\rho^{(1)}-|\tilde{h}^{(1)}|^2}
	{S_n^{(1)}},N^{(1)}\right] \nonumber \\
  & = & \textrm{CDF}\left[\frac{\rho^{(1)}}{S_n^{(1)}}
	-\left(\frac{\rho^{(2)}}{S_n^{(2)}}-N^{(2)}\right)
	\frac{S_n^{(2)}}{S_n^{(1)}}\frac{N^{(1)}\Delta T^{(1)}}
	{N^{(2)}\Delta T^{(2)}},N^{(1)}\right] \; .
\end{eqnarray}
Now, for any choice of $\Delta T^{(1)}$, $\Delta T^{(2)}$, etc., the
thresholds $\rho^{(1)}$ and $\rho^{(2)}$ are completely constrained by
our choices of final confidence level $\alpha$ and false dismissal
probability $D$.  The false dismissal probability is fixed at $D=0.01$
in our optimization; this is an acceptably low level, meaning that
only one signal in a hundred is expected to be lost in this type of
search.

\subsection{Computational costs}
\label{ss:computational-costs}

The computational cost $C^{(1)}$ of the first stage of the search
follows the same formula as for a simple non-hierarchical search,
that is
\begin{eqnarray}
C^{(1)} &=&
6f_{\mathrm{max}} N^{(1)}\Delta T^{(1)} N^{(1)}_{pc}\,
[\log_2(2f_{\mathrm{max}}\Delta
T^{(1)}) \nonumber \\
&& \ \ + 0.5 +
N^{(1)}_{pf}(N^{(1)}-1)/(6N^{(1)}N^{(1)}_{pc})]\; .
\label{eq:hierarchical-cost-1}
\end{eqnarray}
For each of the $F^{(1)}$ first-stage triggers, the second stage
requires $N_{pc}^{(2)}/N_{pf}^{(1)}$ (minimum 1) coarse grid
corrections (each involving $N^{(2)}$ FFT's of length $\Delta
T^{(2)}$), along with $N_{pf}^{(2)}/N_{pf}^{(1)}$ (minimum 1)
frequency shifts and spectrum additions.   Each of the coarse grid
corrections requires the usual $2f_{\mathrm{max}}N^{(2)}\Delta
T^{(2)}[3\log_2(2f_{\mathrm{max}}\Delta T^{(2)})+0.5]$ floating-point
operations.   The incoherent frequency shifts and spectrum additions
require only $2(N^{(2)}-1)\Delta T^{(2)}/\Delta T^{(1)}$~floating
point operations since the frequency correction and power summation
need only be 
applied over a bandwidth of $\sim2$ first-pass frequency
bins.  The total cost of the second pass is therefore:
\begin{figure}[tb]
\vbox{
\vbox{
\psfig{file=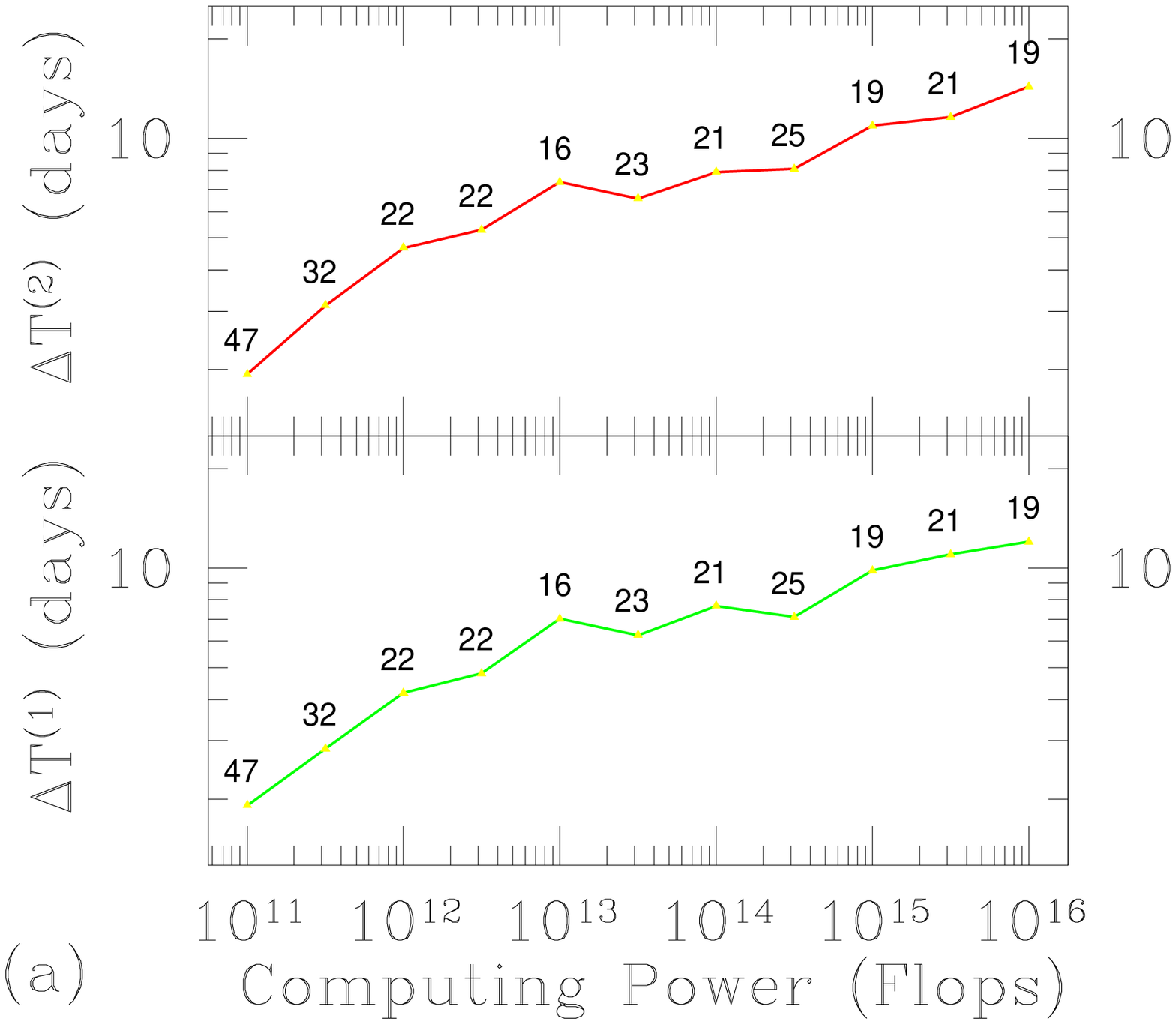,width=9cm,bbllx=0in,%
bblly=1.8in,bburx=8.5in,bbury=8.5in}\vskip -0.2in
\psfig{file=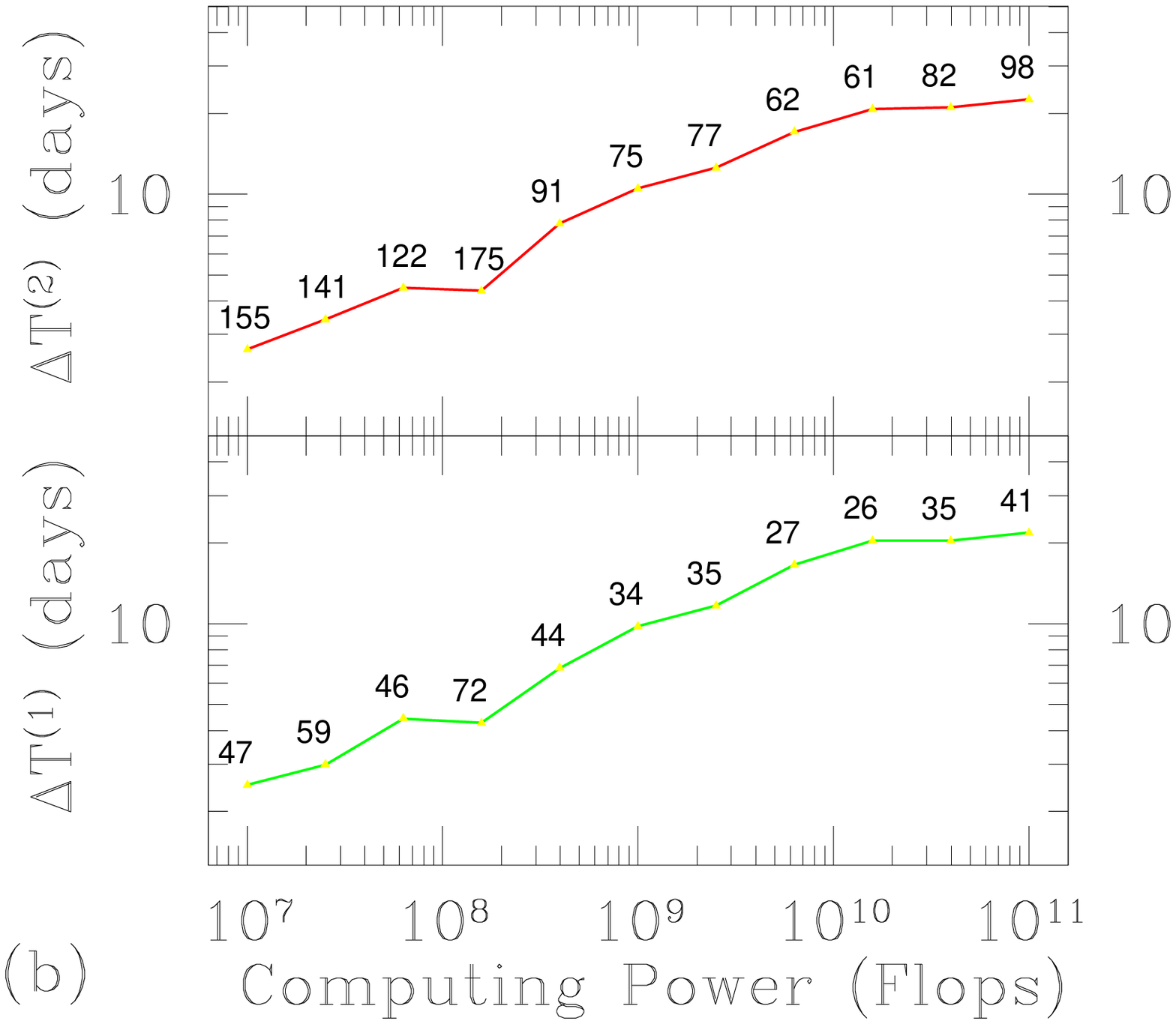,width=9cm,bbllx=0in,%
bblly=1.8in,bburx=8.5in,bbury=8.5in}
}
\caption{\label{fig:stacked-hierarchical-directed} The optimum stack
length $\Delta T$ (thick solid line) and number of stacks $N$ (numbers
along the solid line) for the first and second stages of directed,
hierarchical searches.  For numerical convenience the maximal
projected mismatch was chosen in advance to be $\mu_{\mathrm{max}} =
0.3$, and the number of stacks in the first and second stage are the
same.  The two panels correspond to the following cases: (a) The
situation encountered when searching for periodic sources having
gravitational wave frequencies up to $1000\mbox{\rm Hz}$, with minimum
spindown ages $\tau_{\rm \protect\scriptsize min}=40\mbox{\rm yr}$.
(b) The equivalent results for gravitational wave frequencies up to
$200\mbox{\rm Hz}$, with minimum spindown ages $\tau_{\rm
\protect\scriptsize min}=10^3\mbox{\rm yr}$.  Both cases assume a
threshold which gives an overall statistical significance of $99\%$ to
a detection (although the results are insensitive to the precise
value).  }}
\end{figure}\noindent
\begin{eqnarray}
C^{(2)} &=& \frac{2F^{(1)}N^{(2)}\Delta T^{(2)}
N_{pc}^{(2)}}{N_{pf}^{(1)}}
	\biggl\{3 f_{\mathrm{max}}
	[\log_2(2f_{\mathrm{max}}\Delta T^{(2)})
	\nonumber \\	&& \ \  +0.5] +
	\frac{N^{(2)}_{pf}(N^{(2)}-1)}{N^{(2)} N^{(2)}_{pc} \Delta T^{(1)}}
	\biggr\} \; .
\label{eq:hierarchical-cost-2}
\end{eqnarray}

We require that data analysis proceed at the rate of data acquisition.
Since the amount of data used in the second-stage of the search will
generally be greater than that used in the first, we require that the
analysis be completed in $N^{(2)}\Delta T^{(2)}$ seconds.  Thus the
computational power is given by
\begin{equation}
\label{eq:hierarchical-power}
P=(C^{(1)}+C^{(2)})/N^{(2)}\Delta T^{(2)}\; .
\end{equation}

Our final sensitivity $\Theta$ is given by Eq.~(\ref{eq:sensitivity}),
using the observation time, mismatch level, and threshold of the
\emph{second} stage of the search.  Optimization then consists of
maximizing this function over the six parameters $\Delta T^{(1)}$,
$\Delta T^{(2)}$, $\mu^{(1)}$, $\mu^{(2)}$, $N^{(1)}$, and $N^{(2)}$,
given constraint Eq.~(\ref{eq:hierarchical-power}) for specified
$\alpha$, $D$, and $P$.

\section{Hierarchical search with stacking}
\label{s:hierarchical-search-with-stacking}

It turns out that the optimization described in the previous section
is only weakly sensitive to the parameters $\mu^{(1)}$ and
$\mu^{(2)}$; that is, even if we choose values for $\mu^{(1)}$ and
$\mu^{(2)}$ quite different from the optimal ones, we can recover
nearly all of the sensitivity by adjusting the other parameters for
the same computational power $P$.  In particular, if we arbitrarily
fix $\mu^{(1)}=\mu^{(2)}=0.3$ and re-optimize, we obtain sensitivities
within 20\% of the optimal.

This becomes very useful when we consider the generalized two-stage
hierarchical search \emph{with} stacking.  Normally this would involve
optimizing over six variables ($\mu^{(1),(2)}$, $N^{(1),(2)}$, and
$\Delta T^{(1),(2)}$) with one constraint on $P$.  However, by
assuming that we can continue to set $\mu^{(1)}=\mu^{(2)}=0.3$ with
minimal loss of sensitivity, we can reduce our degrees of freedom back
down to four minus one constraint.

\begin{figure}[tb]
\vbox{
\vbox{
\psfig{file=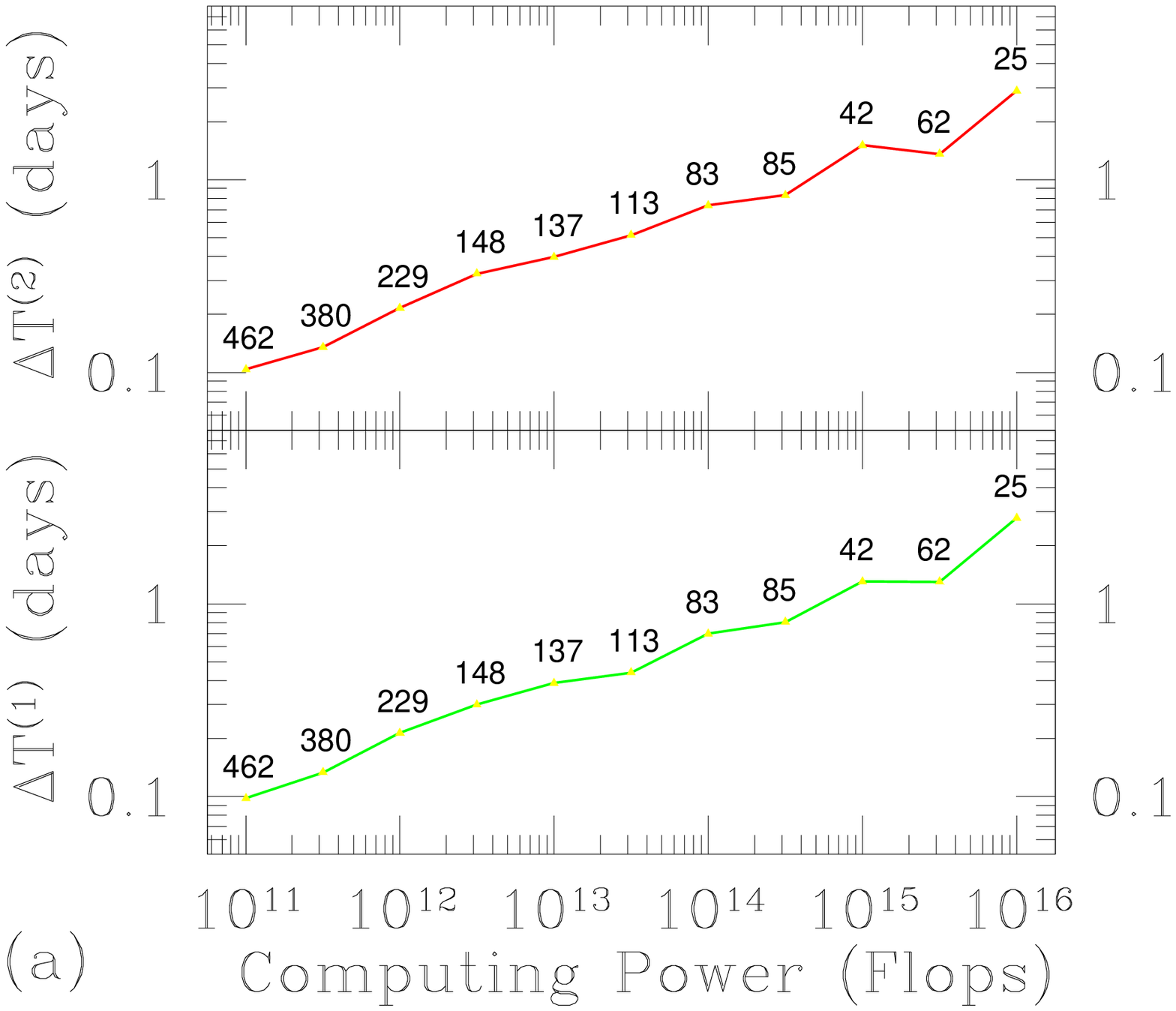,width=9cm,bbllx=0in,%
bblly=1.8in,bburx=8.5in,bbury=8.5in}\vskip -0.2in
\psfig{file=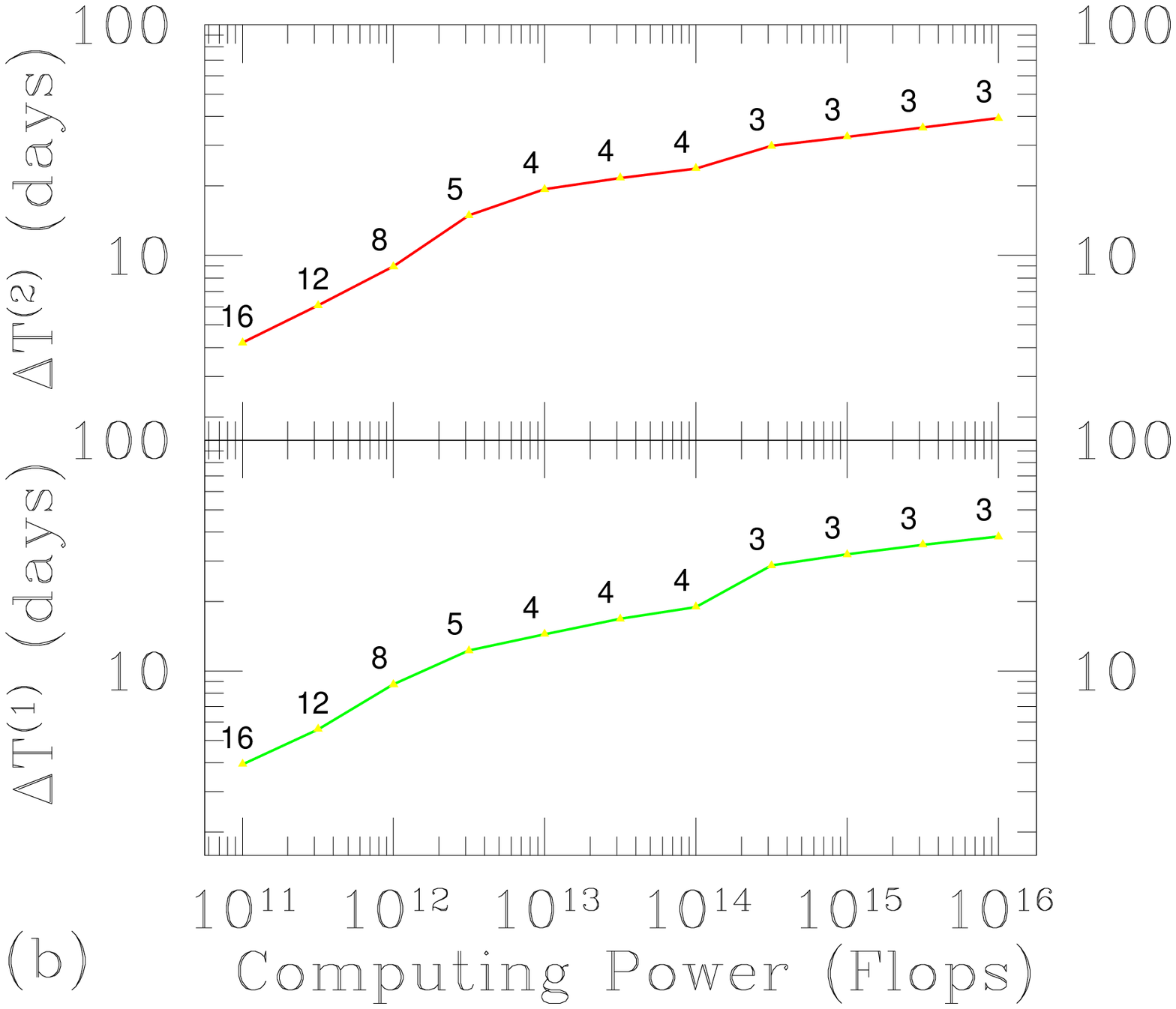,width=9cm,bbllx=0in,%
bblly=1.8in,bburx=8.5in,bbury=8.5in}
}
\caption{\label{fig:stacked-hierarchical-all-sky}
Same as in Fig.~\ref{fig:stacked-hierarchical-directed}, but for an
all-sky search.
}}
\end{figure}

The results of this optimization for our four canonical example
searches are given in Figs.~\ref{fig:stacked-hierarchical-directed}
and \ref{fig:stacked-hierarchical-all-sky}.  We have again chosen a
final confidence level $\alpha=0.99$ and a false dismissal probability
of $D=0.01$.  Figures~\ref{fig:stacked-hierarchical-directed}(a) and
(b) show the results for a directed search for old, slow
($\tau_{\mathrm{min}}$=1\,000~yr, $f_{\mathrm{max}}$=200~Hz) and young,
fast
($\tau_{\mathrm{min}}$=40~yr, $f_{\mathrm{max}}$=1\,000~Hz) pulsars,
respectively.
Figures~\ref{fig:stacked-hierarchical-all-sky}(a) and (b) show the
results for an all-sky search for the same two classes of source.  The
optimal sensitivities achieved by these searches are summarized in
Fig.~\ref{fig:stacked-sensitivities}(b) in the Introduction.

\section{Specialized searches}
\label{s:specialized-searches}

The strongest sources of continuous gravitational waves are
likely to be the most difficult to detect since the frequency of the
waves will be changing significantly as the source radiates angular
momentum.  As we have seen in the previous sections, an all sky search
for these sources is unlikely to achieve the desired sensitivity with
available computational resources.  To reach better sensitivity
levels, it will be useful to consider targeted searches for specific
types of source.  In this section, we consider three such searches:
(i) neutron stars in the galactic core as an example of a limited area
sky survey, (ii) newborn neutron stars triggered on optically observed
extra-galactic supernovae, and (iii) low mass X-ray binary systems
such as Sco~X-1.

\subsection{Galactic core pulsars}
\label{ss:galactic-core-pulsars}

Area surveys of the sky will certainly begin with the region most
likely to hold a large number of nearby sources.  Based on population
models of radio pulsars in our Galaxy~\cite{Curran_S:1995}, there
should be many rapidly rotating neutron stars in the galactic bulge.
As an example of a limited area search, we therefore consider the
optimal strategy for searching an area of 0.004~steradians about the
galactic core, for sources with frequencies $f\leq 500$~Hz and spindown
ages $\tau\ge 100$~yr.  The choice of a 0.004~steradian search is
arbitrary; it includes the entire molecular cloud complex at the core
of the galaxy ($\sim300$~pc radius at a distance of $\sim8.5$~kpc).

It is easy to include a correction factor, to allow for this limited
area, in our calculation of the number of patches by reducing the
ranges of the integral over $\cal P$ in Eq.~(\ref{eq:n-patches}).
Given the approximations in Sec.~\ref{ss:sky-search}, this
amounts to reducing $N_p$ in Eq.~(\ref{eq:np-sky}) by
\begin{equation}
\label{eq:galactic-reduction}
	0.97 \times \left(\frac{0.004}{4\pi}\right) \; ,
\end{equation}
where the multiplicative factor $0.97$ is the correction for the
difference in functional form between the mismatch metric and the
angular area metric $d\Omega^2=\sin^2\theta\,d\theta d\phi$ in the
direction of the galactic center (i.e. $-28.9^\circ$ declination).

\begin{figure}[tb]
\vbox{
\psfig{file=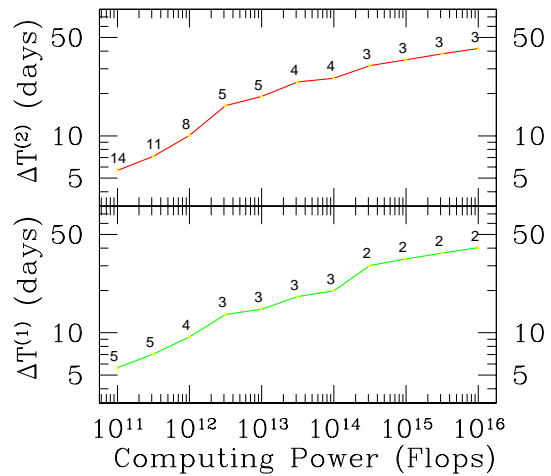,width=9cm,bbllx=0in,%
bblly=2.0in,bburx=8.5in,bbury=8.25in}
\caption{\label{fig:stacked-hierarchical-galactic-centre}
The optimum stack length $\Delta T$ (thick solid line) and number of
stacks $N$ (numbers along the solid line) for the first and second
stages of an hierarchical search for pulsars located in a sky region
of $0.004$ steradians about the Galactic center, with $\tau\ge 100$~yr
and $f\le 500$~Hz.  For numerical convenience the maximal projected
mismatch was chosen in advance to be $\mu_{\mathrm{max}} = 0.3$, and
the number of stacks in the first and second stage are the same.  A
threshold is chosen which gives an overall statistical significance of
$99\%$ to a detection (although the results are insensitive to the
precise value). } }
\end{figure}

The optimal choices of $N^{(1),(2)}$ and $\Delta T^{(1),(2)}$ for a
hierarchical stacked search are shown in
Fig.~\ref{fig:stacked-hierarchical-galactic-centre} as a function of
available computing power; the relative sensitivity of this search is
shown in Fig.~\ref{fig:targeted-sensitivities} of the Introduction.

We note from Eq.~(3.6) of Paper~I that gravitational waves from
rapidly rotating neutron stars might be expected to have a
characteristic amplitude of
\begin{equation}
\label{eq:pulsar-amplitude}
h_c = 2.3\times 10^{-25} \frac{\epsilon}{10^{-5}}
	\frac{I_{zz}}{10^{45}\mathrm{ g cm}^2}
	\frac{8.5\mathrm{ kpc}}{r}
	\left(\frac{f}{500\mathrm{ Hz}}\right)^2 \; ,
\end{equation}
where $\epsilon=(I_{xx}-I_{yy})/I_{zz}$ is the non-axisymmetric
strain, $I_{ij}$ is the moment of inertia tensor, $r$ is the distance
to the source, $f$ is the gravitational wave frequency, and $h_c$ has
been averaged over the detector responses to various source
inclinations~\cite{Thorne_K:1987}.  Theoretical estimates of the
strength of the crystalline neutron star crust suggest that it can
support static deformations of up to $\epsilon\sim10^{-5}$, though
most neutron stars probably support smaller deformations.  From
Figs.~\ref{fig:h_3peryear} and~\ref{fig:targeted-sensitivities}, we
see that 1~Tflops of computing power should allow us to detect
pulsars with strains as small as $\epsilon\sim 5\times 10^{-6}$ at
$8.5\>\textrm{kpc}$ using enhanced LIGO detectors.

\subsection{Newborn neutron stars}
\label{ss:newborn-neutron-stars}

Several recent
papers~\cite{Andersson_N:1998,Friedman_J:1997,Lindblom_L:1998} have
indicated that newly-formed fast-spinning neutron stars may be copious
emitters of gravitational radiation.  If the newborn neutron star is
rotating sufficiently fast, its $r$-modes (axial-vector current
oscillations whose restoring force is the Coriolis force) are unstable
to gravitational radiation reaction.  As the star cools, viscous
interactions eventually damp the modes in isolated neutron stars.
Numerical studies~\cite{Owen_B:1998} indicate that neutron stars which
are born with rotational frequencies above several hundred Hz will
radiate away most of their angular momentum in the form of
gravitational waves during their first year of life.  Estimates of the
viscous time scales, and the superfluid transition temperature,
suggest that the $r$-modes are stabilized when the star cools below
$\sim10^9$~K and are rotating at $\sim 100$--$200$~Hz.  During the
evolutionary phase when most of the angular momentum is lost, the
amplitude and spindown time scale are expected to be
\begin{eqnarray}
\label{eq:r-mode-h}
h_c &=& 1.2\times 10^{-24}\sqrt{\kappa}
	\left(\frac{f}{1\mathrm{ kHz}}\right)^3
	\left(\frac{20\mathrm{ Mpc}}{r}\right) \\
\label{eq:r-mode-tau}
\tau &\approx& \frac{580\mathrm{ s}}{\kappa}
	\left(\frac{1\mathrm{ kHz}}{f}\right)^6
	\approx {6t} \; .
\end{eqnarray}
These estimates are based on Eqs.~(4.9) and~(5.13) in
Ref.~\cite{Owen_B:1998}. (We note that the ``characteristic
amplitude'' used in Ref.~\cite{Owen_B:1998} is appropriate to estimate
the strength of burst sources, and is different from our $h_c$.) Here
$\kappa$ is a dimensionless constant of order unity; it parameterizes
our ignorance of the non-linear evolution of the $r$-mode instability.
The distance to the neutron star is $r$, and $t$ is the actual
age of the star.  Figure~\ref{fig:h_3peryear} shows $h_c$ as a
function of frequency with $\kappa=1$ at distances $r=2$~Mpc and
$20$~Mpc.

\begin{figure}[tb]
\vbox{
\psfig{file=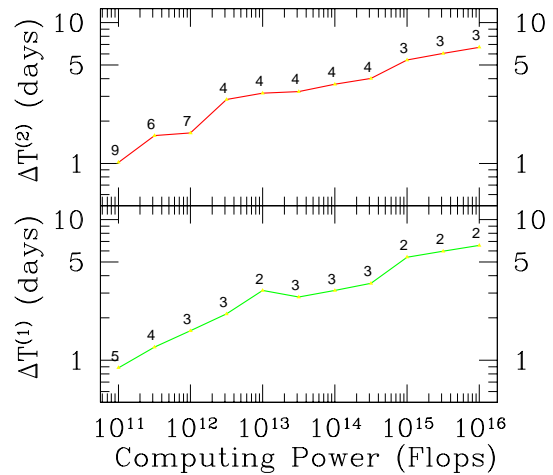,width=9cm,bbllx=0in,%
bblly=2.0in,bburx=8.5in,bbury=8.25in}
\caption{\label{fig:stacked-hierarchical-r-modes}
The optimum stack length $\Delta T$ (thick solid line) and number of
stacks $N$ (numbers along the solid line) for
the first and second stages of a hierarchical search for newborn
neutron stars spinning down due to an active $r$-mode instability.
We assume that a supernova has been identified and accurately located
on the sky,  so this is a directed search for an object with
$\tau_{\protect\mathrm{min}}=1$~yr,  and $f\leq 200$~Hz.  For numerical
convenience the maximal projected mismatch was chosen in advance to be
$\mu_{\protect\mathrm{max}} = 0.3$, and the number of stacks in the first
and
second stage are the same.  A threshold is chosen which gives
an overall statistical significance of $99\%$ to a detection
(although the results are insensitive to the precise value).
}}
\end{figure}

Sources outside our Galaxy are potentially detectable due to the high
gravitational-luminosity of a newborn neutron star with an active
$r$-mode instability.  Nevertheless, it is a significant challenge to
develop a feasible search strategy for these signals since the
frequency evolves on such short time scales (compared to those
considered above).  One approach is to perform directed searches on
optically observed supernova explosions.  Although some supernovae may
not be optically visible, and this instability may not operate in all
newborn neutron stars, the computational benefits of targeting
supernovae are substantial (if not essential).  Based on the estimates
in Ref.~\cite{Owen_B:1998}, most of the signal to noise is accumulated
during the final stages of spindown.  With limited computational
resources, it seems best to limit the directed searches to frequencies
$\lesssim 200$~Hz, when the spindown time scale is $\sim 1$~yr.
Figure~\ref{fig:stacked-hierarchical-r-modes} shows the optimal search
criteria in a hierarchical stacked search for neutron stars aged two
months or older; the upper frequency cutoff is $f_{\mathrm{max}}=200$~Hz
and
the minimum spindown timescale is $\tau_{\mathrm{min}}=1$~yr.  The
sensitivities achievable in a search are shown in
Fig.~\ref{fig:targeted-sensitivities} of the Introduction.

Figure~\ref{fig:targeted-sensitivities} shows that $1$~Tflops of
computing power will \emph{not} suffice to detect newborn neutron
stars as far away as the Virgo cluster ($\approx 20$~Mpc); however,
such sources will be marginally detectable within $\sim8$~Mpc by
enhanced LIGO detectors.  The NBG catalog~\cite{Tully_R:1988} lists
165 galaxies within this distance (assuming a Hubble expansion of
75~km/s/Mpc, retarded by the Virgo cluster).  From the Hubble types
and luminosities of these galaxies, and the supernova event rates
in~\cite{vandenBergh_S:1994}, we estimate a total supernova rate of
$\sim0.6$ per year in this volume, of which $\sim10$\% would be of
type~Ia, $\sim20$\% of type~Ib or Ic, and $\sim70$\% of type~II.  (We
note that the total rate is consistent with values given in
Ref.~\cite{Talbot_R:1976}.)  At present, it is not known what fraction
of these will produce neutron stars with unstable $r$-modes.
\vskip -.3in
\subsection{X-ray binaries}
\label{ss:xray-binaries}

A low-mass x-ray binary (LMXB) is a neutron star orbiting around a
stellar companion from which it accretes matter.  The accretion process
deposits both energy and angular momentum onto the neutron star.  The
energy is radiated away as x-rays, while the angular momentum
spins the star up.  Bildsten~\cite{Bildsten_L:1998} has suggested that
the accretion could create non-axisymmetric temperature gradients in
the star, resulting in a substantial mass quadrupole and gravitational
wave emission.  The star spins
up until the gravitational waves are strong enough to radiate away the
angular momentum at the same rate as it is accreting; according to
Bildsten's estimates the equilibrium occurs at a gravitational-wave
frequency $\sim 500$~Hz.  The characteristic gravitational-wave
amplitudes from these sources would be
\begin{eqnarray}
h_c &\agt& 4\times10^{-27}\left(\frac{R}{10\mathrm{km}}\right)^{3/4}
	\left(\frac{M}{1.4M_\odot}\right)^{-1/4} \nonumber \\
	&& \ \ \times
	\left(\frac{F}{10^{-8}\mathrm{erg\,cm}^{-2}\mathrm{s}^{-1}}
	\right)^{1/2}
	\left(\frac{f}{600\,\mathrm{Hz}}\right)^{-1/2} \; ,
\label{eq:xray-h}
\end{eqnarray}
where $R$ and $M$ are the radius and mass of the neutron star, and $F$
is the observed x-ray flux at the Earth.

The amplitude of the gravitational waves from these sources make them
excellent candidates for targeted searches.  If the source is an x-ray
binary pulsar---an accreting neutron star whose rotation is observable
in radio waves---then one can apply the exact phase correction deduced
from the radio timing data to optimally detect the gravitational
waves.  (In this process, one must assume a relationship between the
gravitational-wave and radio pulsation frequencies.)  Unfortunately,
radio pulsations have not been detected from the rapidly rotating
neutron stars in all LMXB's (i.e.  neutron stars which rotate hundreds
of times a second).  In the absence of direct radio observations,
estimates of the neutron-star rotation rates are obtained from
high-frequency periodic, or quasi-periodic, oscillations in the x-ray
output during Type~I x-ray bursts. (See Ref.~\cite{vanderKlis_M:1998} for a
summary.)  But this does not provide precise timing data for a
coherent phase correction.  To detect gravitational waves from these
sources, one must search over the parameter space of Doppler
modulations due to the neutron-star orbit around its companion, and
fluctuations in the gravitational-wave frequency due to variable
accretion rates.  The Doppler effects of the gravitational-wave
detector's motion can be computed exactly, because the sky position of
the source is known.

In most cases, the orbital period of an x-ray binary can be deduced
from periodicity in its x-ray or optical light curve.  In some cases,
the radial component of the orbital velocity can be computed by
observing an optical emission line from the accretion disk, as was
done with the bright x-ray binary Sco~X-1~\cite{Cowley_A:1975}.  Such
observations do not determine the phase modulation of the
gravitational-wave signal with sufficient precision to make the search
trivial; however, they do substantially constrain the parameter space
of modulations.

In this subsection, we consider a directed search for an x-ray binary
in which the orbital parameters are known up to an uncertainty $\delta
v$ in the radial velocity $v_r$ of the neutron star, and an
uncertainty $\delta\phi$ in the orbital phase.  It is assumed that
long-term photometric observations of the source can give the orbital
period $P$ to sufficient precision that we need not search over it
explicitly.  We therefore parameterize the phase modulations as
follows:
\begin{equation}
\label{eq:xray-phase}
\phi(t;\mbox{\boldmath${\lambda}$\unboldmath}) =
2\pi f_0 \left(t+\frac{v_1 P}{2\pi c}\cos2\pi t/P
	+ \frac{v_2 P}{2\pi c}\sin2\pi t/P\right) \; ,
\end{equation}
where $\mbox{\boldmath${\lambda}$\unboldmath}=(f_0,v_1,v_2)$ are our
search parameters, the gravitational-wave frequency $f_0$ is
constrained to be $\leq f_{\mathrm{max}}$, and the pair $(v_1,v_2)$ is
constrained to lie within an annular arc of radius $v_r$, width
$\delta v$, and arc angle $\delta\phi$.

Applying the formalism developed in section~\ref{s:mismatch} to this
problem gives essentially the same result as for a sky search over
Earth-rotation-induced Doppler modulations if one converts time units
by the ratio $P/\mathrm{day}$.  In the case of the Earth's rotation,
a search over sky positions $\hat{n}$ corresponds to a search over an
area $\pi v_{\mathrm{rot}}^2\cos^2(\lambda)$ in the equatorial components
of
the source's velocity relative to the detector, whereas in the case of
a binary orbit, the search is over a coordinate area $v_r\delta
v\delta\phi$.  So we can simply multiply equation~(\ref{eq:np-sky}) by
the ratio of these coordinate areas to obtain the number of grid
points $N_p$ in the parameter space:
\begin{eqnarray}
\label{eq:np-xray}
N_p &\approx& \frac{(f_{\mathrm{max}}P)^2}{2\mu_{\mathrm{max}}}
	\frac{v_r\delta v\delta\phi}{c^2}
	\left(A^{-2}+B^{-2}+C^{-2}\right)^{-1/2} \; , \\
\label{eq:a-xray}
A &=& 0.5 \; , \\
\label{eq:b-xray}
B &=& 1.6 (\Delta T/P)^2 \; , \\
\label{eq:c-xray}
C &=& 6.4 (\Delta T/P)^5 N^3 \; .
\end{eqnarray}

Accounting for the intrinsic phase variations of the spinning neutron
star itself is problematic.  Baykal and \"Ogelman~\cite{Baykal_A:1993}
showed empirically that x-ray pulsar frequencies could be well-modeled
as a random walk, plus possibly a secular spin up for
rapidly-accreting systems.  Over a time $t$, the angular rotation rate
would undergo excursions up to $\Delta\Omega=\sqrt{St}$, where
$S\sim10^{-17}\mathrm{rad}^2\mathrm{s}^{-3} (L_x/10^{37}
\mathrm{erg\,s}^{-1})$ and $L_x$ is the x-ray luminosity of the
source.  For the sources of interest to us, accretion proceeds at or
near the Eddington rate ($L_x\sim3\times10^{38}$~erg/s), and the
gravitational-wave frequency is $f=\Omega/\pi$, so we expect frequency
drifts $\sim2\times10^{-6}\mathrm{Hz}\sqrt{t/\mathrm{days}}$.  If we
require that the frequency drifts by less than one Fourier bin $\Delta
f\leq1/t$ during a coherent observation, the observing time $t$ must
satisfy
\begin{equation}
\label{eq:t-max}
t \leq 3.2\>\mathrm{days.}
\end{equation}
\begin{figure}[tb]
\vbox{
\psfig{file=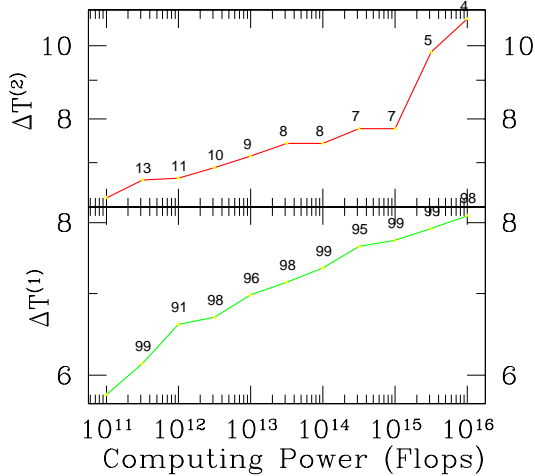,width=9cm,bbllx=0in,%
bblly=2.0in,bburx=8.5in,bbury=8.25in}
\caption{\label{fig:stacked-hierarchical-xray} The optimum stack
length $\Delta T$ (thick solid line) and number of stacks $N$ (numbers
along the solid line) for the first and second stages of a
hierarchical search for sources of continuous gravitational waves
which are in binary systems.  We assume the orbit is characterized by
two orthogonal velocity parameters which are known to within a total
error of $17(\mathrm{km/s})^2$, and that the frequency $f\leq500$~Hz
is experiencing a random walk typical of Eddington-rate accretion.
For numerical convenience the maximal projected mismatch was chosen in
advance to be $\mu_{\mathrm{max}} = 0.3$.  A threshold is chosen which
gives an overall statistical significance of $99\%$ to a detection
(although the results are insensitive to the precise value).  }
}
\end{figure}\vskip -0.1in

This type of random walk cannot be modeled as a low-order polynomial
in time.  Nevertheless, the stack-slide technique is well suited to
search for these sources since corrections for the stochastic changes
in frequency can be applied by shifting the stacks by +1, 0, or $-$1
frequency bins when required.  In a search using $N$ stacks, each of
length $\Delta T$, this kind of correction would be applied after a
time $t$ such that
\begin{equation}
2\times10^{-6}\mathrm{Hz}\sqrt{t/\mathrm{days}} = 1/\Delta T\; .
\label{eq:correction-time}
\end{equation}
The number of times that these corrections must be applied is then
$N\Delta T/t$, and the number of distinct frequency evolutions traced
out in this procedure is $3^{N\Delta T/t}$.  (Monte-Carlo simulations
of stacked FFTs of signals undergoing random walks in frequency have
shown that one can increase $t$ by up to a factor of 4, i.e.\ allowing
drifts of up to $\pm2$~frequency bins, while incurring only $\sim20\%$
losses in the final summed power.)  We have not yet studied in detail
how this combines with the mismatches generated from other
demodulations, or how to search over all demodulations together in an
optimal way.  For now we assume a factor of $3^{0.01N(\Delta
T/\mathrm{day})^3}$ extra points in our search mesh for mismatches of
$\mu_{\mathrm{max}}\simeq0.3$.

As an example, we consider a search for gravitational waves from the
neutron star in Sco~X-1.  This system has an orbital period
$P=0.787313\pm0.000001$~days~\cite{Gottlieb_E:1975}, a radial orbital
velocity amplitude of $v_r=58.2\pm3.0$~km/s, an orbital phase known to
$\pm0.10$~radians~\cite{Cowley_A:1975}, and an inferred
gravitational-wave frequency
$f_{\mathrm{max}}\approx500$~Hz~\cite{Bildsten_L:1998}.  We note that the
uncertainty in $P$ is basically negligible over the $<3.2$~day
coherent integrations expected.  The remaining uncertainties give:
\begin{eqnarray}
\label{eq:np-scorpius}
N_p &=& \frac{10^5}{\mu_{\mathrm{max}}}3^{0.01N(\Delta T/\mathrm{day})^3}
	\left(A^{-2}+B^{-2}+C^{-2}\right)^{-1/2} \; , \\
\label{eq:a-scorpius}
A   &=& 0.5 \; , \\
\label{eq:b-scorpius}
B   &=& 2.6(\Delta T/\mathrm{day})^2 \; , \\
\label{eq:c-scorpius}
C   &=& 21N^3(\Delta T/\mathrm{day})^5 \; .
\end{eqnarray}
where it is understood that $\Delta T\leq3.2$~days, in order for the
random-walk stack-slide corrections to achieve maximum sensitivity.

Figure~\ref{fig:stacked-hierarchical-xray} shows the optimal search
criteria for a hierarchical stacked search for the Sco~X-1 pulsar
under these assumptions.  The sensitivities achievable in such a
search are shown in Figure~\ref{fig:targeted-sensitivities} of the
Introduction.  We see that $1$~Tflops of computing power may be
sufficient to detect this source using enhanced LIGO detectors if it
is radiating most of the accreting angular momentum as gravitational
waves.  The sensitivity to these sources might be enhanced by a factor
$\sim 5$--$10$ if the interferometer is operated in a signal-recycled,
narrow-band configuration during the search.

\mbox{\ }
\vskip -0.45in
\mbox{\ }
\section{Future directions}

\vskip -0.1in

We have presented in this manuscript the rudiments of a search
algorithm for sources of continuous gravitational waves.  The next
step is to implement some variant of these schemes for a simple test
search; a good starting point would be a directed, acceleration (a
single spindown parameter) search of a nearby supernova remnant.  This
will be sufficient to test every stage of the technique, and to assess
the accuracy of the computational estimates presented here.  It will
also allow direct comparison with other approaches such as the
line-tracking method that is being explored by Papa~\cite{Papa_M:1998}.

Further theoretical work is required to determine the parameter space
which should be searched, especially in the case of active $r$-mode
instabilities and radiating neutron stars in LMXB's.  Finally, it
would be worthwhile to consider in detail what advantage, if any, can
be gained by using data from multiple interferometers at the initial
detection stage of a search for continuous gravitational waves.

\mbox{\ }
\vskip -0.45in
\mbox{\ }

\section*{Acknowledgments}
\vskip -0.1in

This work was supported by NSF grant PHY94-24337.  PRB is
supported by NSF grant PHY94-07194 at the ITP, and he is grateful to
the Sherman Fairchild Foundation for financial support while at
California Institute of Technology.  We especially thank Kip Thorne
for his help and encouragement throughout this work, and Stuart
Anderson for many illuminating discussions.  This work has also
benefited from interactions with Bruce Allen, Curt Cutler, Jolien
Creighton, Sam Finn, Scott Hughes, Andrzej Krolak, Ben Owen, Tom
Prince, Bernard Schutz, and Alan Wiseman.

\onecolumn
\appendix

\section{Patch number formulae}
\label{s:patch-number-formulae}

The approximate formulae given in Eqs.~(\ref{eq:g1})--(\ref{eq:g3})
are valid when $N\gg4$.  General expressions for the $G_s$ can be
derived by setting $\vec{x}=\vec{0}$ in Eqs.~(\ref{eq:gw-phase}) and
(\ref{eq:tb}), and using Eqs.~(\ref{eq:delta-phase}),
(\ref{eq:metric-stacked-total}), (\ref{eq:metric-stacked-individual}),
(\ref{eq:average-def}), (\ref{eq:projected-metric}), and
(\ref{eq:n-patches}):
\begin{eqnarray}
\label{eq:analytic-g1}
G_1(N) &=& \frac{\pi}{6\sqrt{5}}\sqrt{5N^2-4} \\
\label{eq:analytic-g2}
G_2(N) &=& \frac{\pi^2}{180\sqrt{105}}\sqrt{175N^6-840N^4+1100N^2
	-432} \\
\label{eq:analytic-g3}
G_3(N) &=& \frac{\pi^3}{75600\sqrt{105}}\sqrt{3675N^{12}-58800N^{10}
	+363160N^8-1053360N^6+1484336N^4-987840N^2+248832} \; .
\end{eqnarray}
\noindent In Eq.~(\ref{eq:n-patches}) we approximate the metric as having
constant determinant and evaluate it at the point of zero spindown;
this introduces small errors of order $(\mbox{[}N\Delta
T\mbox{]}/\tau)^2$.

Equations~(\ref{eq:ms-sky})--(\ref{eq:c-sky}) for the number of sky
patches ignoring spindown and orbital motions provide an empirical fit
to a numerical evaluation of the metric determinant.  The determinant
was found to have an approximate functional dependence
$\sqrt{|\gamma_{ij}|}\sim |\sin2\theta|$ with corrections of order
$v/c\approx10^{-4}$.  Assuming this dependence to be exact,
Eqs.~(\ref{eq:projected-metric}) and~(\ref{eq:n-patches}) give
\begin{equation}
\label{eq:analytic-ms-sky}
{\cal M}_s = \frac{2/\pi}{4\mu_{\mathrm{max}}/(s+2)}
	\sqrt{\det\left|g_{ij}-\frac{g_{0i}g_{0j}}{g_{00}}
	\right|_{{}^{f=f_{\mathrm{max}}}_{\theta=\pi/2}}}\; .
\end{equation}
Here $g_{\alpha\beta}$ is the mismatch metric, defined in
Eqs.~(\ref{eq:metric-stacked-total})--(\ref{eq:average-def}), computed
using only the Earth-rotation-induced Doppler modulation.  Since the
Earth's rotation is a simple circular motion, and since we are
evaluating the metric at a single point in parameter space, we can
carry out the integrals in Eq.~(\ref{eq:average-def}) analytically, to
obtain
\begin{equation}
\label{eq:analytic-metric}
g_{\alpha\beta}=b_{\alpha\beta}-\frac{1}{N}\sum_{k=1}^N
	a_{k\alpha}a_{k\beta} \; ,
\end{equation}
where
\begin{eqnarray}
\label{eq:analytic-metric-a0}
a_{k0} &=& \left(k-\frac{1}{2}\right)\Delta\Phi +
	\frac{v/c}{\Delta\Phi}\left\{
	\sin(k\Delta\Phi)-\sin(\mbox{[}k-1\mbox{]}\Delta\Phi)\right\} \\
\label{eq:analytic-metric-a1}
a_{k1} &=& \frac{f_{\mathrm{max}}v/c}{\Delta\Phi}\left\{
	\sin(k\Delta\Phi)-\sin(\mbox{[}k-1\mbox{]}\Delta\Phi)\right\} \\
\label{eq:analytic-metric-a2}
a_{k2} &=& \frac{f_{\mathrm{max}}v/c}{\Delta\Phi}\left\{
	\cos(\mbox{[}k-1\mbox{]}\Delta\Phi)-\cos(k\Delta\Phi)\right\} \\
\label{eq:analytic-metric-b00}
b_{00} &=& \frac{1}{12\Phi}\left\{-24+\frac{v}{c}
	+6\left(\frac{v}{c}\right)^2\Phi+4\Phi^3+24\frac{v}{c}\cos\Phi
	+24\frac{v}{c}\Phi\sin\Phi
	+3\left(\frac{v}{c}\right)^2\sin2\Phi\right\} \\
\label{eq:analytic-metric-b11}
b_{11} &=& \frac{f_{\mathrm{max}}^2(v/c)^2}{4\Phi}(2\Phi+\sin2\Phi) \\
\label{eq:analytic-metric-b22}
b_{22} &=& \frac{f_{\mathrm{max}}^2(v/c)^2}{4\Phi}(2\Phi-\sin2\Phi) \\
\label{eq:analytic-metric-b01}
b_{01} &=& b_{10} = \frac{f_{\mathrm{max}}v}{4\Phi}
	\left\{-4+2\frac{v}{c}\Phi+4\cos\Phi+4\Phi\sin\Phi
	+\frac{v}{c}\sin2\Phi\right\} \\
\label{eq:analytic-metric-b02}
b_{02} &=& b_{20} = \frac{f_{\mathrm{max}}v/c}{2\Phi}\left\{-2\Phi\cos\Phi
	+\sin\Phi\left(2+\frac{v}{c}\sin\Phi\right)\right\} \\
\label{eq:analytic-metric-b12}
b_{12} &=& b_{21} = \frac{f_{\mathrm{max}}^2(v/c)^2}{2\Phi}\sin^2\Phi \; .
\end{eqnarray}
Here $\Phi=2\pi T_b/(1\,\mathrm{day})$ is the total angle over which the Earth
rotates during the observation, $\Delta\Phi=\Phi/N$ is the angle
rotated during each stretch of the data, and $v$ is the maximum radial
velocity relative to the detector at latitude $\lambda=45^\circ$ of a
point at a polar angle $\theta=\pi/4$ on the sky, that is
\begin{equation}
\label{eq:v-sky}
v = \frac{2\pi R_{\mathrm{earth}}\cos\lambda\sin\theta}{1\,\mathrm{day}} \; .
\end{equation}

\narrowtext
\twocolumn
\section{Resampling error}
\label{app:resampling-error}

In this paper, we have assumed that coherent phase corrections are
achieved through stroboscopic resampling: a demodulated time
coordinate $t_b\mbox{[}t]$ is constructed, and the data stream $h(t)$ is
sampled at equal intervals in $t_b$ at the Nyquist rate for the
highest frequency signal present,
$f_{\mathrm{Nyquist}}=2f_{\mathrm{max}}$.  However, since the data
stream is initially sampled at some finite rate
$f_s=Rf_{\mathrm{Nyquist}}$ (where $R$ is the \emph{oversampling
factor}), this can introduce errors: in general, there will not be a
data point exactly at a given value of $t_b$, so the nearest (in time)
datum must be substituted.  Consequently, there will be residual phase
errors $\Delta\Phi(t)\in[-\pi/2R,\pi/2R)$ caused by rounding to the
nearest datum even if one chooses a phase model whose frequency and
modulation parameters exactly match the signal. The phase of the
resampled signal drifts until the timing error is $\geq1/2f_s$, at
which point one corrects the phase by sampling an adjacent datum,
which shifts in time by $1/f_s$.  These residual phase errors reduce
the Fourier amplitude of the signal by a fraction
\begin{equation}
\label{eq:resample-sum}
F = \left| \frac{1}{N}\sum_{k=1}^N
e^{i\Delta\Phi(k/f_{\mathrm{Nyquist}})} \right| \; ,
\end{equation}
where $N=f_{\mathrm{Nyquist}}\Delta T$ is the total number of points in
the
resampled data stream.

The uncorrected signal will in general drift by many radians in phase,
which is the reason why we must apply phase corrections in the first
place.  This means that $\Delta\Phi(t)$ will sweep through the range
$[-\pi/2R,\pi/2R)$ many times over the course of the observation.  So,
regardless of the precise form of the phase evolution, we expect
$\Delta\Phi(k/f_{\mathrm{Nyquist}})$ to be essentially evenly distributed
over this interval.  Thus, replacing the sum with an expectation
integral, we have
\begin{equation}
\label{eq:resample-integral}
F \approx \left| \frac{R}{\pi}\int_{-\pi/2R}^{\pi/2R}e^{i\Phi}\,d\Phi
	\right| = \frac{\sin(\pi/2R)}{\pi/2R} \; .
\end{equation}
The fractional losses in amplitude $1-F$ for a few values of the
oversampling factor $R$ are given in Table~\ref{table:resample}.

\vbox{\begin{table}
\vbox{
\caption{\label{table:resample}
The percentage reduction $(1-F)$ in amplitude of a signal as a
function of the oversampling factor $R$.   The LIGO interferometers will
collect data at $f_s = 16\;384$Hz,  so that the data will be oversampled
by $R\geq 4$ compared to the maximum gravitational wave frequency that
we expect on physical grounds.   In fact,  it seems more likely that
$R \simeq 8$ for real signals.}
\begin{tabular}{r@{=\quad}c@{\quad}c@{\quad}c@{\quad}c@{\quad}%
c@{\quad}c@{\quad}c@{\quad}c}
 $R$ &  2    &  3   &  4   &  5   &  6   &  7   &  8 \\
$1-F$&10.0\% & 4.5\%& 2.6\%& 1.6\%& 1.1\%& 0.8\%& 0.6\%
\end{tabular}
}
\end{table}}

The highest gravitational-wave frequencies we consider are 1000~Hz,
requiring a resampling rate of $f_{\mathrm{Nyquist}}=2000$~Hz.  Since
LIGO will acquire data at a rate of $16\;384$Hz, corresponding to an
oversampling factor of $R>8$, we have a maximum signal loss due to
resampling of $1-F=0.6\%$.  Resampling errors will increase if the
number of data samples is reduced by some factor \emph{before}
phase-correcting.



\begin{thebibliography}{99}

\bibitem{Brady_P:1998prd}
P.~R. Brady, T. Creighton, C. Cutler, and B.~F. Schutz, Phys. Rev. {\bf
D57},
  2101  (1998).

\bibitem{Livas_J:1987}
J.~C. Livas, Ph.D. thesis, Massachusetts Institute of Technology, 1987.

\bibitem{Jones_G:1995}
G.~S. Jones, Ph.D. thesis, University of Whales, 1995.

\bibitem{Niebauer_T:1993}
T.~M. Niebauer {\it et~al.}, Phys. Rev. D {\bf 47},  3106  (1993).

\bibitem{Chandrasekhar_S:1970}
S. Chandrasekhar, Phys. Rev. Let. {\bf 24},  611  (1970).

\bibitem{Friedman_J:1978}
J.~L. Friedman and B.~F. Schutz, Ap. J. {\bf 222},  281  (1978).

\bibitem{Bonazzola_S:1996}
S. Bonazzola and E. Gourgoulhon, Astron. Astr. {\bf 312},  675  (1996).

\bibitem{Zimmermann_M:1980}
M. Zimmermann, Phys. Rev. D. {\bf 21},  891  (1980).

\bibitem{Zimmermann_M:1979}
M. Zimmermann and E. Szedenits, Jr., Phys. Rev. D. {\bf 20},  351  (1979).

\bibitem{Wagoner_R:1984}
R.~V. Wagoner, Astrophys. J. {\bf 278},  345  (1984).

\bibitem{Galtsov_D:1984}
D.~V. Gal'tsov, V.~P. Tsvetkov, and A.~N. Tsirulev, Sov. Phys. ---JETP
{\bf
  59},  472  (1984).

\bibitem{New_K:1995}
K.~C.~B. New, G. Chanmugam, W.~W. Johnson, and J.~E. Tohline, Ap. J. {\bf
450},
   757  (1995).

\bibitem{Krolak_A:1998}
A. Krolak,  Searching data for periodic signals, gr-qc/9803055.

\bibitem{Jaranowski_P:1998a}
P. Jaranowski, A. Krolak, and B.~F. Schutz, Phys. Rev. {\bf D58},
063001 (1998), gr-qc/9804014.

\bibitem{Jaranowski_P:1998b}
P. Jaranowski and A. Krolak,  Data analysis of gravitational wave signals
from spinning neutron stars. 2. Accuracy of estimation of parameters,
  gr-qc/9809046.

\bibitem{Papa_M:1998}
M. A. Papa, B. F. Schutz, S. Frasca and P. Astone, Detection of 
continuous gravitational wave signals: pattern tracking with the 
Hough transform,  to appear in Proceedings of the LISA Symposium
(1998);   M.A. Papa, P. Astone, S.Frasca and B.F. Schutz,  Searching
for continuous waves by line identification,  to appear in the
proceedings of Gravitational Wave Data Analysis Workshop (1997);
M.~A. Papa,  private communication.

\bibitem{Shapiro_S:1984}
S.~L. Shapiro, S.~A. Teukolsky, and I. Wasserman, Ap. J. {\bf 272},  702
  (1984).

\bibitem{Friedman_J:1984}
J.~L. Friedman, J.~R. Ipser, and L. Parker, Nature {\bf 312},  255
(1984).

\bibitem{Kulkarni_S:1992}
S.~R. Kulkarni, Phil. Trans. R. Soc. Lond. A {\bf 341},  77  (1992).

\bibitem{Andersson_N:1998}
N. Andersson, Astrophys. J. {\bf 502},  708 (1998), gr-qc/9706075.

\bibitem{Friedman_J:1997}
J.~L. Friedman and S.~M. Morsink, Astrophys. J. {\bf 502}, 714 (1998),
gr-qc/9706073.

\bibitem{Lindblom_L:1998}
L. Lindblom, B.~J. Owen, and S.~M. Morsink, Phys. Rev. Lett. {\bf 80},
4843
  (1998).

\bibitem{Owen_B:1998}
B.~J. Owen {\it et~al.}, Phys. Rev. {\bf D58},  084020 (1998),
gr-qc/9804044.

\bibitem{Bildsten_L:1998}
L. Bildsten, Astrophys. J. {\bf 501},  L89 1998, astro-ph/9804325.

\bibitem{Schutz_B:1991}
B.~F. Schutz,  in {\em The Detection of Gravitational Waves}, edited by
D.~G.
  Blair (Cambridge University Press, Cambridge, 1991), Chap.~16, pp.\
406--451.


\bibitem{Anderson_S:1993}
The method of stacking power-spectra has been used by radio
astronomers in deep searches for millisecond pulsars,  although all
coreections were applied to the data stream via resampling and not
sliding the spectra.  For more information on the implementation,  see
S.~B.~Anderson, Ph.D. thesis,  California Institute of Technology, 1993.

\bibitem{Thorne_K:1987}
K.~S. Thorne,  in {\em Three hundred years of gravitation}, edited by
S.~W.
  Hawking and W. Israel (Cambridge University Press, Cambridge, 1987),
Chap.~9,
  pp.\ 330--458.

\bibitem{Schutz_B:GWDAW}
B.~F. Schutz,  Sources of radiation from neutron stars, gr-qc/9802020;
and,  talk given at Gravitational Wave Data Analysis Workshop,  MIT
(1996).

\bibitem{Owen_B:1996}
B. Owen, Phys. Rev. D. {\bf 53},  6749  (1996).

\bibitem{Curran_S:1995}
S.~J. Curran and D.~R. Lorimer, Mon. Not. R. Astron. Soc. {\bf 276},  347
  (1995).

\bibitem{Tully_R:1988}
R.~B. Tully, {\em Nearby Galaxy Catalog} (Cambridge University Press,
  Cambridge, 1988).

\bibitem{vandenBergh_S:1994}
S. van~den Bergh and R.~D. McClure, Ap. J. {\bf 425},  205  (1994).

\bibitem{Talbot_R:1976}
R. Talbot, Ap. J. {\bf 205},  535  (1976).

\bibitem{vanderKlis_M:1998}
M. van~der Klis,  in {\em The Many Faces of Neutron Stars}, edited by A.
Alpar,
  L. Buccheri, and J. van Paradijs (Kluwer, Dordrecht, 1998 (in press)).

\bibitem{Cowley_A:1975}
A.~P. Cowley and D. Crampton, Ap. J. Letters {\bf 201},  L65  (1975).

\bibitem{Baykal_A:1993}
A. Baykal and H. \"Ogelman, Astron. Astrophys. {\bf 267},  119  (1993).

\bibitem{Gottlieb_E:1975}
E.~W. Gottlieb, E.~L. Wright, and W. Liller, Ap. J. Letters {\bf 195},
L33
  (1975).

\end{thebibliography}
\end{document}